%% file: main.tex
\documentclass[sigconf]{acmart}

\pdfoutput=1

\usepackage[labelformat=simple,skip=0pt]{subcaption}
\usepackage[textfont=md,skip=0pt]{caption}

\usepackage{physics, tabularx, hyphenat, graphicx, booktabs}

\newcommand{\systemname}{Quantum Belief Propagation}
\newcommand{\systemnames}{Quantum Belief Propagation's}
\newcommand{\systemacronym}{QBP}
\newcommand{\systemacronyms}{QBP's}
\newcommand{\sysembedder}{QGEM}

\newcommand{\parahead}[1]{\vspace{2pt plus 0pt minus 2pt}\noindent{\bfseries #1}}
\newcommand{\parabreak}{\vspace*{1.00ex minus 0.25ex}\noindent}

\renewcommand{\paragraph}[1]{\vspace{2pt plus 0pt minus 2pt}\noindent{\bfseries #1}}

\copyrightyear{2020}
\acmYear{2020}
\setcopyright{acmcopyright}\acmConference[MobiCom '20]{The 26th Annual International Conference on Mobile Computing and Networking}{September 21--25, 2020}{London, United Kingdom}
\acmBooktitle{The 26th Annual International Conference on Mobile Computing and Networking (MobiCom '20), September 21--25, 2020, London, United Kingdom}
\acmPrice{15.00}
\acmDOI{10.1145/3372224.3419207}
\acmISBN{978-1-4503-7085-1/20/09}

\begin{document}

\title{Towards Quantum Belief Propagation for LDPC Decoding in Wireless Networks}

\author{Srikar Kasi and Kyle Jamieson}
\email{{skasi, kylej} @cs.princeton.edu}
\affiliation{Department of Computer Science, Princeton University}
\email{}

\begin{abstract}
\input{abstract.tex}

\end{abstract}




\maketitle

\input{intro.tex}

\input{primer-ldpc.tex}
\input{limitations.tex}

\input{primer-qa.tex}
\input{design.tex}
\input{implementation.tex}
\input{evaluation.tex}

\input{related.tex}
\input{discussion.tex}
\input{conclusion.tex}
\input{acknowledgements}

\clearpage
\bibliographystyle{ACM-Reference-Format}
\bibliography{reference}

\end{document}

%% file: abstract.tex
We present Quantum Belief Propagation (\systemacronym{}), a Quantum Annealing (QA) based decoder design for Low Density Parity Check (LDPC) error control codes, which have found many useful applications in Wi-Fi, satellite communications, mobile cellular systems, and data storage systems. \systemacronym{} reduces the LDPC decoding to a discrete optimization problem, then embeds that reduced design onto quantum annealing hardware. QBP's embedding design can support LDPC codes of block length up to 420 bits on real state-of-the-art QA hardware with 2,048 qubits. We evaluate performance on real quantum annealer hardware, performing sensitivity analyses on a variety of parameter settings. Our design achieves a bit error rate of $10^{-8}$ in 20 $\mu$s and a 1,500 byte frame error rate of $10^{-6}$ in 50 $\mu$s at SNR 9~dB over a Gaussian noise wireless channel. Further experiments measure performance over real-world wireless channels, requiring 30~$\mu$s to achieve a 1,500 byte 99.99\% frame delivery rate at SNR 15-20~dB. \systemacronym{} achieves a performance improvement over an FPGA based soft belief propagation LDPC decoder, by reaching a bit error rate of $10^{-8}$ and a frame error rate of $10^{-6}$ at an SNR 2.5--3.5~dB lower. In terms of limitations, QBP currently cannot realize practical protocol-sized (\textit{e.g.,} Wi-Fi, WiMax) LDPC codes on current QA processors. Our further studies in this work present future cost, throughput, and QA hardware trend considerations.

%% file: intro.tex
\section{Introduction}
\label{s:intro}

As the design of mobile cellular wireless networks continues to
evolve, time\hyp{}critical baseband processing functionality from the base stations at the very edge of the wireless network is being shifted and aggregated into more centralized locations (\textit{e.g.,} Cloud\fshyp{}Centralized\hyp{}RAN \cite{5429914, 10.1145/2505906.2511048, cloudran-ieee15}) or even small edge datacenters. A key component of mobile cellular baseband
processing is the error correction code, a construct that adds parity
bit information to the data transmission in order to correct the bit
errors that interference and the vagaries of the wireless channel
inevitably introduce into the data. In particular LDPC codes, first
introduced by Gallager\cite{gallager1962low} in 1962 but (with few
exceptions \cite{zyablov1975estimation, tanner1981recursive,
margulis1982explicit}) mostly ignored until the work of McKay \emph{et
al.}\ in the late 90s \cite{mackay1999good}, have approached the
Shannon rate limit \cite{shannon1948mathematical}. Along with Turbo codes \cite{397441}, LDPC codes stand out today because of their exceptional error correcting capability even close to capacity, but their decoding comprises a significant fraction of the processing requirements for a mobile cellular base station. LDPC codes are considered for inclusion in the 5G New Radio traffic channel \cite{etsi2018138}, the DVB-S2 standard for satellite communications \cite{morello2006dvb}, and deep space communications \cite{book2020radio, book2014erasure}. LDPC codes are also currently utilized in
the most recent revisions of the 802.11 Wi-Fi protocol family
\cite{6178212}. Given the dominance of LDPC codes in today's wireless networks,
the search for computationally efficient decoders and
their ASIC\fshyp{}FPGA realization is underway.

\paragraph{Background: Quantum Annealing.}
This paper notes exciting new developments
in the field of computer architecture hold the potential to efficiently decode LDPC codes: recently, \emph{quantum annealer} (QA) machines previously
only hypothesized \cite{AQC, kadowaki1998quantum} have been commercialized, and are now
available for use by researchers.  QA machines are
specialized, analog computers that solve NP\hyp{}complete and
NP\hyp{}hard optimization problems in their Ising specification \cite{bian2010ising} on current hardware, with future
potential for substantial speedups over conventional computing
\cite{10.1145/2482767.2482797}.  They are comprised of an array of physical devices, each
representing a single \emph{physical qubit} (quantum bit), that can
take on a continuum of values, unlike classical information bits,
which can only take on binary values.  The user of the QA inputs a set
of desired pairwise \emph{constraints} between individual qubits
(\emph{i.e.}, a slight preference that two particular qubits should
differ, and\fshyp{}or a strong preference that two particular qubits
should be identical) and preferences that each individual qubit should
take a particular classical value (0 or 1) in the solution the machine
outputs.  The QA then considers the entire set of constraints as a
large optimization problem that is typically expressed as a quadratic
polynomial of binary variables \cite{kadowaki1998quantum, 10.3389/fphy.2014.00005}.  A
multitude of quantum annealing \textit{trials} comprises a single
machine \emph{run}, with each anneal trial resulting in a potentially
different solution to the problem: a set of classical output bits, one
per qubit, that best fits the user\hyp{}supplied constraints on that
particular trial.

\paragraph{Quantum-Inspired and Hybrid Algorithms.} The growing interest in quantum computing has recently led to the emergence of several physics-based \textbf{quantum-inspired} algorithms (QIA) \cite{han2002quantum, aramon2019physics, katzgraber2006feedback, 1905.10415} and quantum-classical \textbf{hybrid} algorithms (QCH) \cite{tran2016hybrid, mcclean2016theory, sweke2019stochastic, irie2020hybrid}. QIA can be used to simulate quantum phenomena such as \textit{superposition} and \textit{entanglement} on classical hardware \cite{montiel2019quantum}, where widely practiced QIA approaches (\textit{e.g.,} \emph{digital annealing}  \cite{aramon2019physics, 9045100}) have solved combinatorial optimization problems with as many as 8,192 problem variables \cite{9045100}. QCH algorithms broadly operate on a hybrid workflow between classical search heuristics and quantum queries, providing ways to use noisy intermediate-scale quantum computers \cite{preskill2018quantum} for optimizing problems with as many as 10,000 variables \cite{hybrid}. In this work, while we demonstrate a quantum annealing based LDPC decoder approach by realizing a small 700 variable problem, we also note that implementation of the same ideas using QIA and QCH methods is also a promising possibility.

\parabreak
This paper presents \emph{\systemname{}}
(\emph{\systemacronym{}}), a new uplink LDPC decoder that takes a new
look at error control decoding, from the fresh perspective of the
quantum annealer.  \systemacronym{} is a novel way to design an LDPC
decoder that sets aside traditional \textit{belief propagation} (BP) decoding, instead \emph{reduces} the first principles of the LDPC code construction in a highly\hyp{}efficient way directly onto the physical
grid of qubits present in the QA we use in this study, the D-Wave 2000-qubit (DW2Q) quantum adiabatic optimizer machine, taking into account the
practical, real\hyp{}world physical qubit interconnections. We have empirically evaluated \systemacronym{} on the real DW2Q QA hardware. Results on the real\hyp{}world quantum annealer
show that \systemacronym{} achieves a bit error rate of $10^{-8}$ in 20~$\mu$s
and a 1,500 byte frame error rate of $10^{-6}$ in 50~$\mu$s at
signal-to-noise ratio of 9~dB over a Gaussian noise channel.~In comparison with FPGA-based soft BP LDPC decoders, QBP achieves the same $10^{-8}$ bit error rate and $10^{-6}$ frame error rate at an SNR 2.5--3.5 dB lower, even when the classical decoder is allowed a very large number of iterations~(100). Currently, QBP cannot realize practical protocol-sized LDPC codes on state-of-the-art QA processors with 2,048 qubits. Our further studies present limitations and predicted future of QA (\S\ref{s:discussion}).

%% file: primer-ldpc.tex
\section{Primer: LDPC codes}
\label{sec:2}\label{s:primer-ldpc}

\begin{figure}
\centering
\includegraphics[width=0.73\linewidth]{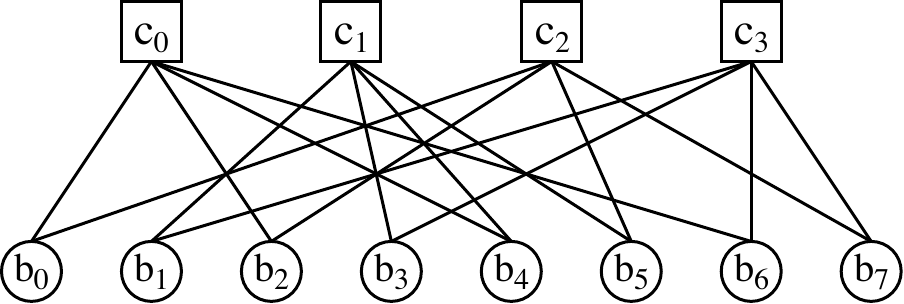}
\caption{A \emph{Tanner Graph} of an example LDPC code.}
\label{fig:tanner}\label{f:tanner}
\end{figure}

A binary (\emph{N}, \emph{K}) LDPC code is a linear block code described functionally by a sparse parity check matrix \cite{gallager1962low, mackay1999good}. It is said to be a ($d_{b}$, $d_{c}$)-regular code if every bit node participates in $d_{c}$ checks, and every check has $d_{b}$ bits that together constitute a \textit{check constraint}. This section describes the conventional encoding and decoding schemes of LDPC codes.
Let \textbf{H} = $[h_{ij}]_{M\times N}$ be the LDPC parity check matrix. Each row in \textbf{H} represents a check node constraint whereas each column indicates which check constraint a bit node participates in.
In the Tanner graph \cite{tanner1981recursive} of Figure~\ref{fig:tanner}, the 
nodes labeled $c_{i}$ are check nodes and those labeled $b_{i}$
are bit nodes, and a value of 1 
at $h_{mn}$ $\in$ \textbf{H} represents an edge between $c_{m}$ and 
$b_{n}$. Code \emph{girth}, the length of the shortest cycle
in the Tanner graph, is a crucial measure, as a low girth affects 
the independence of information 
exchanged between check and bit nodes, diminishing the code's 
performance \cite{lu2006structured,1023274, chilappagari2008girth}.

\paragraph{LDPC Encoder.} 
Let \emph{u} be a message of length \emph{K}. 
The overall encoding process is summarized as follows:

\begin{enumerate}
\item Convert \textbf{H} into \textit{augmented form} $[\textbf{P}|\textbf{I}_{N-K}]$ 
by Gauss\hyp{}Jordan elimination.  (Here, \textbf{P} is obtained in the conversion process and
\textbf{I} is the identity matrix.)

\item Construct a generator matrix \textbf{G} as $[\textbf{I}_{K}|\textbf{P}^T]$.

\item The encoded message \emph{c} is constructed as \emph{c} = \emph{u}\textbf{G}.
\end{enumerate}
This way of encoding ensures that the modulo two bit\hyp{}sum at every check node is zero \cite{gallager1962low}.

\paragraph{LDPC Decoder.} 
\label{s:primer-ldpc:decoder}
We describe the BP-based \textit{min-sum} algorithm \cite{zhao2005implementation}. Let \emph{y} be received information, \emph{N($c_{m}$)} the set of bit nodes participating in check constraint \emph{$c_{m}$}, and \emph{M($b_{n}$)} the set of check nodes connected to bit node \emph{$b_{n}$}.

\parahead{Initialization.} Initialize all the bit nodes with their respective \emph{a priori} log-likelihood ratios (LLRs) as:
\setlength{\abovedisplayskip}{5pt}
\setlength{\belowdisplayskip}{5pt}
\begin{equation}
LLR_{b_{n}}(x_{n}) = \log \Bigg(\frac {Pr\big(b_{n} = 0\textbf{|} \emph{y}\big)} {Pr\big(b_{n} = 1\textbf{|} \emph{y}\big)}\Bigg) \hspace{10pt} \forall b_{n} \in N(c_{m})
\end{equation}
 
\parahead{Step 1.} For every combination \{$(m,n)$ $|$ $h_{mn}$ = $1$\}, initialize messages sent to check $c_{m}$ from bit $b_{n}$ $\in$ \emph{N($c_{m}$)} as:
\setlength{\abovedisplayskip}{5pt}
\setlength{\belowdisplayskip}{5pt}
\begin{equation}
\textstyle{Z_{b_{n}\to c_{m}} (x_{n}) = LLR_{b_{n}}(x_{n})}
\end{equation}

\parahead{Step 2.} Every check node $c_{m}$ then updates the message to be sent back, w.r.t every $b_{n}$ $\in$ \emph{N($c_{m}$)} as:

\begin{equation}
\textstyle{Z_{c_{m}\to b_{n}} (x_{n})=\displaystyle\prod_{b_{n'} \in N(c_{m}) \setminus b_{n}}\text{sgn}(Z_{b_{n'}\to c_{m}})\cdot \text{min}|Z_{b_{n'}\to c_{m}}|}
\end{equation}

% _{ b_{n'} \in N(c_{m}) \setminus b_{n}}
\parahead{Step 3.} Each bit node $b_n$ now updates the message to send back, w.r.t every $c_m \in M(b_n)$ as:
\setlength{\abovedisplayskip}{5pt}
\setlength{\belowdisplayskip}{5pt}
\begin{equation}
\textstyle{Z_{b_{n}\to c_{m}} (x_{n}) = LLR_{b_{n}}(x_{n}) + \displaystyle\sum_{c_{m'} \in \emph{M($b_{n}$)}\emph{$\setminus c_{m}$}} Z_{c_{m'} \to b_{n}}(x_{n})}
\end{equation}
To decode, each bit node computes:
\setlength{\abovedisplayskip}{5pt}
\setlength{\belowdisplayskip}{5pt}
\begin{equation}
\label{eq:4}
\textstyle{Z_{b_{n}} (x_{n}) = LLR_{b_{n}}(x_{n}) + \displaystyle\sum_{c_{m} \in \emph{M($b_{n}$)}} Z_{c_{m} \to b_{n}}(x_{n})}.
\end{equation}

\parahead{Decision Step.} After Step 3, quantize $\Hat{\textbf{x}}$ = [$\Hat{x}_{0}$, $\Hat{x}_{1}$, ... , $\Hat{x}_{N-1}$] such that $\Hat{x}_{n}$ = 0 if $Z_{b_{n}} (x_{n})$ $\geqslant$ 0, else $\Hat{x}_{n}$ = 1. $\Hat{\textbf{x}}$ are the decoded bits. If $\Hat{\textbf{x}}$ satisfies the condition enforced at encoding ($\Hat{\textbf{x}}\textbf{H}^T$ = 0), then $\Hat{\textbf{x}}$ is declared as the final decoded message. If it doesn't satisfy this condition, the BP algorithm iterates Steps~1--3 until a satisfactory $\Hat{\textbf{x}}$ is obtained. The decoder terminates at a predetermined threshold number of iterations.

%% file: limitations.tex
\section{Classical BP Decoder Limitations}
\label{sec:3}\label{s:limitations}

The goal of most classical BP LDPC decoders is an
efficient hardware implementation that maximizes throughput, thus driving
a need to minimize data errors.
A variety of architectures for the classical hardware implementation 
of LDPC decoders have been developed \cite{hailes2016survey, hocevar2004reduced, 
4115107}, and in practice, depending on the problem 
of interest and the hardware resource availability, the decoders are implemented either in 
serial, partly\hyp{}parallel, or fully parallel architectures on FPGA/ASIC hardware.  
Although existing decoders do reach 
theoretically supported line speeds of, \emph{e.g.} Wi-Fi \cite{dot11-2012}, 
Wi-MAX \cite{dot16-2012}, and DVB-S/S2 \cite{dvb-s2}, they make throughput compromises, in particular, reducing decoding precision (such as using low precision LLR bitwidth, limiting iterations, or using reduced\hyp{}complexity algorithms \cite{hailes2016survey}).
Therefore, the goal of maximizing throughput requires making the most
efficient trade\hyp{}offs among the following:
\begin{enumerate}
    \item To achieve high throughput, a high degree of {\bf decoding parallelism}
    is required, demanding more resources in the silicon hardware implementation.
    
    \item Accurate decoding results require high {\bf LLR bit 
    precision} (\emph{ca.} $8-10$), along with a precise decoding algorithm, again demanding more hardware resources.
    
    \item The iterative nature of the BP algorithm impedes throughput 
    by requiring {\bf numerous serial iterations} before reaching the best,
    final result.  Thus a trade\hyp{}off between iteration limit 
    and throughput must be made.
\end{enumerate}

\label{s:limitations:bram}
These tradeoffs induce network designers to 
compromise between decoder operation line rate and precision, within the 
available limited silicon hardware resources. \textit{Block RAMs} (BRAMs) are the fundamental array storage resources in FPGAs, where state-of-the-art BRAMs have a read and a write port with independent clocks, implying that 
a single BRAM can perform a maximum of two read\fshyp{}write 
operations in parallel \cite{xilinx, Boncalo17}. Therefore, to realize a high degree of parallelism required in protocol sized LDPC codes, many BRAMs must be used in parallel to access the BP LLRs. Furthermore to meet FPGA device timing constraints, today's dual-ported support for BRAMs limits 
the size of a single data access to 2,048 bits and the number of BRAMs accessible in a single clock cycle to 1,024 \cite{kastner2018parallel, 5447856, xilinx}. This limitation results in the maximum degree of achievable parallelization in 
current top\hyp{}end Xilinx 
FPGAs, which corresponds to a 2,048 (1,024 $\times$ 2) 
LDPC code block length. However, practical block lengths reach up to 1,944 bits in Wi-Fi, 2,304 bits in Wi-MAX, and 64,800 bits in DVB-S2 protocol standards \cite{dot11-2012, dot16-2012, dvb-s2}.

\paragraph{A Xilinx FPGA Resource Study.} Using the Xilinx synthesis tool \textit{Vivado HLS}, we have implemented a min-sum algorithm based decoder for a $\frac{1}{2}$-rate, 
1944 block\hyp{}length, 
(3, 6)\hyp{}regular LDPC code, on the Xilinx Virtex 
Ultrascale 440 (\emph{xcvu440}, the most resourceful Xilinx FPGA) 
with 8\hyp{}bit LLR precision. The resource measurement metric in FPGAs is generalized to a \textit{Configurable Logic Block} \textit{(CLB)}. Each CLB in the Ultrascale architecture contains eight six-input LUTs\footnote{Most recent FPGAs are equipped with six\hyp{}input 
LUTs, which is equivalent to $1.6 \times$ the resources of 
a four\hyp{}input LUT \cite{clb, hailes2016survey}.}, and 16 flip-flops along with arithmetic carry logic and multiplexers \cite{clb}. Our implementation of this fully parallel LDPC decoder 
covers $\approx 72\%$ (229,322$/$316,220) of the CLBs in the device, the upper limit of reliability in terms of resource utilization. Furthermore, our HLS implementation of a (4,8)-regular LDPC code of block length 2048 bits (fully parallel decoder with 8-bit LLR precision) does not fit into that FPGA.

%% file: primer-qa.tex
\section{Primer: Quantum Annealers}
\label{sec:4}\label{s:primer-qa}

\begin{figure}
    \centering
    \includegraphics[viewport= 0 0 91 44,clip,width=0.7\linewidth]{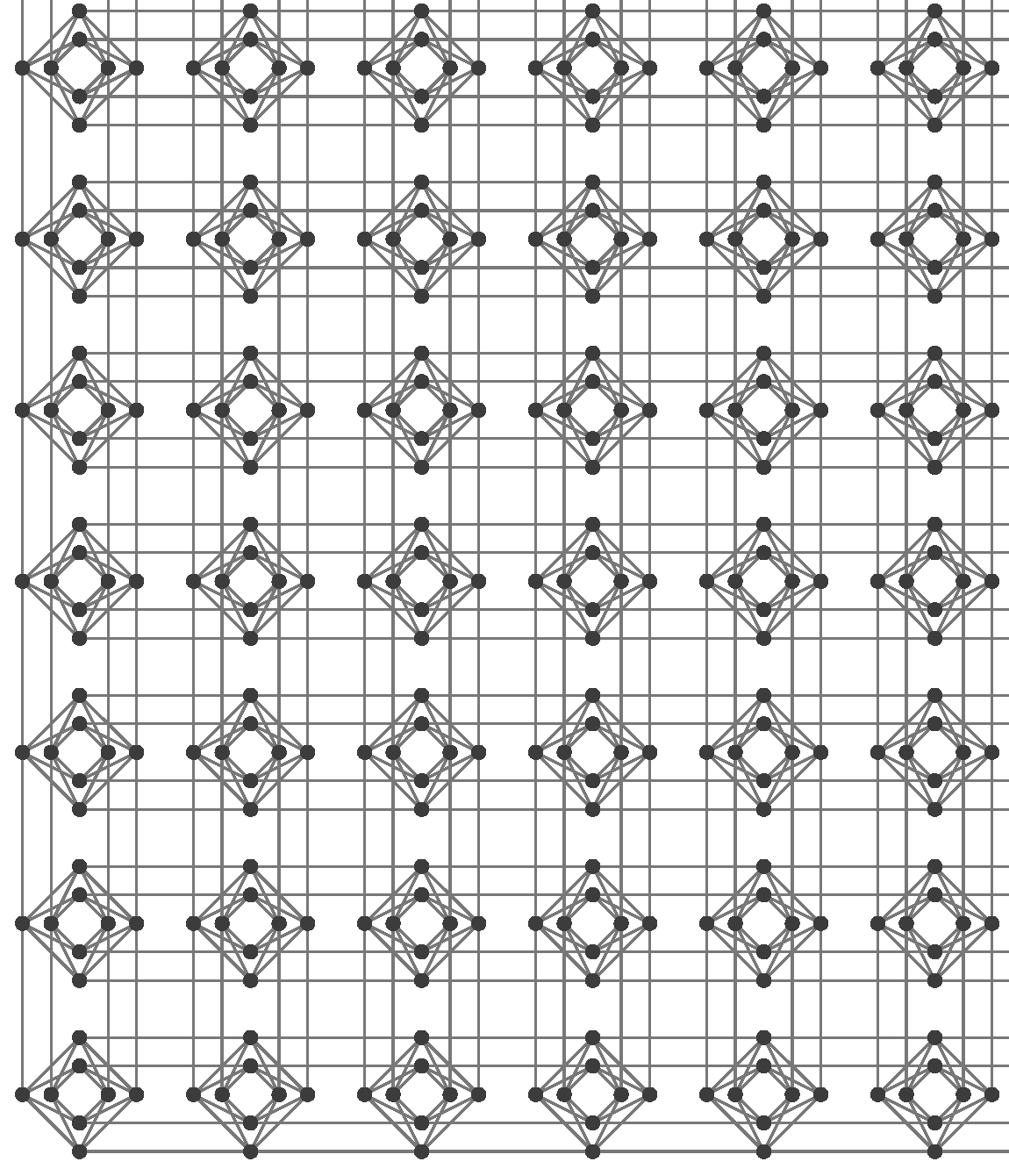}
    \caption{A portion of the Chimera qubit connectivity graph of the DW2Q
    QA, showing qubits (nodes in the figure) grouped by \emph{unit cell}. The edges in the figure are \emph{couplers}.} 
    \label{f:chimera}
\end{figure}

\textit{Quantum Annealing} is a heuristic approach to solve combinatorial 
optimization problems and can be understood at a high level as solving the
same class of problems as the more familiar \emph{simulated annealing}
\cite{10.5555/59580} techniques.
QA takes advantage of the fundamental fact that any process in nature 
seeks a minimum energy state to attain stability. Given a discrete optimization
problem as input, a QA \textit{quantum processor unit} (QPU) internally frames it as an energy minimization 
problem and outputs its ground state as the solution.

\paragraph{Quantum Annealing Fundamentals.} In the QA literature, qubits are classified into two types:\ \emph{physical} and \emph{logical}. A physical qubit is a qubit that is directly available physically on the QA hardware, while a logical qubit is a set of physical qubits. It is often the case that the QA hardware lacks a \textit{coupler} between a particular pair of physical qubits that the user would like to correlate. To construct such a relationship, it is general practice to use intermediate couplers to make several physical qubits behave similarly, as explained below in \S\ref{s:qa-primer:logical}, a process known as \textit{embedding}. The set of these similarly behaving embedded physical qubits is then referred to a \textit{logical qubit}. The process of evolution of quantum bits to settle down at the ground state in the DW2Q QA is called an \textit{anneal}, while the time taken for this evolution is called the \textit{annealing time}. The strength of the preference given to each single qubit to end up in a particular 0 or 1 state is a \textit{bias}, while the strength of each coupler is called \textit{coupler strength}. Moreover the strength of the couplers that are used to make physical qubits behave similarly as in the aforementioned embedding process, are called \textit{JFerros}.

\paragraph{Quantum Annealing Hardware.} The QA processor hardware is a network of interlaced radio-frequency superconducting quantum interference device flux qubits fabricated as an integrated circuit, where the local longitudinal fields (\textit{i.e.,} biases) of the devices are adjustable with an external magnetic field and the interactions (\textit{i.e.,} couplers) between pairs of devices are realized with a tunable magnetic coupling using a programmable on-chip control circuitry \cite{johnson2011quantum, king2018observation}. The interconnection diagram of the DW2Q QA hardware we use in this study is a quasi-planar bi-layer \emph{Chimera} graph. Fig.~\ref{f:chimera} shows a 2$\times$4 portion of the 16$\times$16 QA's Chimera graph:\ each set of eight physical qubits in the figure is called a Chimera \emph{unit cell}, whereas each edge in the figure is a \textit{coupler}.

\paragraph{The Annealing Process.} QA processors simulate systems in the two-dimensional transverse field Ising model described by the time-dependent Hamiltonian:
\begin{align}
    H(s) = - A(s)\sum_i\sigma_i^x + B(s)H_P,\\
    H_P = \sum_i h_{i}\sigma_i^z + \sum_{i<j}J_{ij}\sigma_i^z\sigma_j^z.
\end{align}

where $\sigma_i^{x,z}$ are the Pauli matrices acting on the $i^{th}$ qubit, $h_i$ and $J_{ij}$ are the problem parameters, $s = t/t_a$ where $t$ is the time and $t_a$ is the annealing time. $A(s)$ and $B(s)$ are two monotonic signals such that at the beginning of the anneal (\textit{i.e.,} $t=0$), $A(0) >> B(0) \approx 0$ and at the end of the anneal (\textit{i.e.,} $t=t_a$), $B(1) >> A(1) \approx 0$. The annealing processor initializes every qubit in a \textit{superposition} state $\frac{1}{\sqrt{2}} \left(\ket{0} + \ket{1}\right)$ that has no classical counterpart, then gradually evolves this Hamiltonian from time $t=0$ until $t=t_a$ by introducing quantum fluctuations in a low-temperature environment. The time-dependent evolution of these signals \textit{A} and \textit{B} is essentially the annealing algorithm. During the annealing process, the system ideally stays in the local minima and probabilistically finds the global minimum energy configuration of the problem Hamiltonian $H_P$ at its conclusion \cite{amin2015searching,dwave}.

\subsection{QA Problem Forms}
\label{s:primer-qa:forms}

QA processors can be used to solve the class of \emph{quadratic 
unconstrained binary optimization} (QUBO) problems in their equivalent Ising specification \cite{bian2010ising, 10.1145/3341302.3342072}, which we define 
here. The generalized Ising\fshyp{}QU\-BO form is:
\begin{equation}
E = \sum_{i}h_{i}q_{i} + \sum_{i<j}J_{ij}q_{i}q_{j}.
\label{eqn:ising_qubo}
\end{equation}
Ising form solution variables $\left\{q_i\right\}$ take values in $\left\{-1, +1\right\}$,
and in QUBO form they take values in $\left\{0, 1\right\}$. The linear 
coefficient $h_{i}$ is the bias of $q_i$, whereas the quadratic coefficient $J_{ij}$ is the strength of the coupler between $q_i$ and $q_j$.
Coupler strengths can be used to make the qubits agree or disagree. 
For instance, let us consider an example Ising problem:
\begin{equation}
E = J_{12}q_{1}q_{2},\quad\left(q_{1}, q_{2}\in \{-1, +1\}\right).
\end{equation}
\textbf{Case~I:} $J_{12} = +1$. The energies for qubit states 
$(q_1, q_2)$ = $(-1, -1)$, $(-1, +1)$, $(+1, -1)$, and $(+1, +1)$ 
are $+1$, $-1$, $-1$, and $+1$ respectively. Hence a strong 
\textbf{positive coupler strength} obtains a minimum energy of $-1$
when the two qubits \textbf{are opposites} of each other.

\noindent
\textbf{Case II:}~$J_{12} = -1$. The energies for qubit states $(q_1, q_2)$ = 
$(-1, -1)$, $(-1, +1)$, $(+1, -1)$, and $(+1, +1)$ are $-1$, $+1$, $+1$, 
and $-1$ respectively. Hence a strong \textbf{negative coupler strength}
obtains a minimum energy of $-1$ when the two 
qubits \textbf{agree} with each other.

\subsection{Embedding of Logical Qubits}
\label{s:qa-primer:logical}

\begin{figure}
\centering
\includegraphics[width=0.75\linewidth]{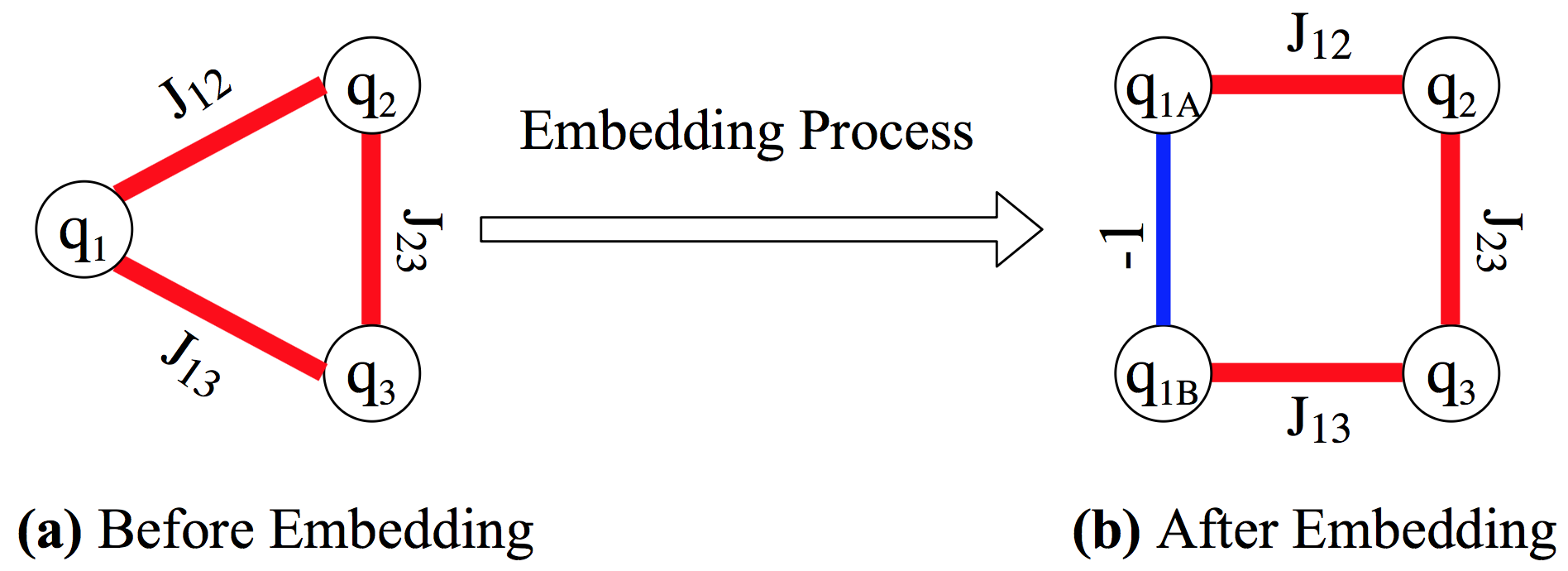}
\caption{The embedding process of Eq.~\ref{fig:emb}, where the logical qubit $q_1$ in (a) is mapped onto two physical qubits $q_{1A}$ and $q_{1B}$ as in (b) with a JFerro of $-1$; here $q_{1A}$ and $q_{1B}$ agree.}
\label{fig:embex}\label{f:embex}
\end{figure}

To visualize the relationship between logical and physical qubits, let us consider another example problem:
\begin{equation}
E = J_{12}q_{1}q_{2} + J_{23}q_{2}q_{3} + J_{13}q_{1}q_{3}.
\label{fig:emb}
\end{equation}
Figure~\ref{f:embex}(a) is the direct graphical representation of this example problem. However, observe that a three\hyp{}node, fully\hyp{}connected graph structure does not exist in the Chimera graph (\emph{cf.} Figure~\ref{f:chimera}). Hence, the standard solution is to \emph{embed} one of the logical qubits into a physical realization consisting of two physical qubits, as Figure~\ref{f:embex}(b) shows, such that we can construct each required edge in Figure~\ref{f:embex}(a). Here, logical qubit $q_1$ is mapped to two physical qubits, $q_{1A}$ and $q_{1B}$ with a JFerro of $-1$ to make $q_{1A}$ and $q_{1B}$ agree with each other.

%% file: design.tex
\section{Design}
\label{sec:5}\label{s:design}

In this section we first detail \systemnames{} 
reduction of the LDPC decoding problem into a quadratic polynomial (QUBO) form (\S\ref{s:design:qubo}), and then present \systemacronyms{} graph
embedding model (QGEM) design on real QA hardware (\S\ref{s:design:embedding}).

\subsection{\systemacronyms{} LDPC to QUBO Reduction}
\label{sec:5.1}\label{s:design:qubo}

Our QUBO reduction (\S\ref{s:obj_fcn}) is a linear combination of two functions we have created:\ {\bf (1)}~an \emph{LDPC satisfier} function (\S\ref{s:sat_fcn}), and {\bf (2)}~a \emph{distance} function (\S\ref{s:dist_fcn}). During an anneal, the LDPC satisfier function leaves all the valid LDPC codewords in the zero energy level while raising the energy of all invalid codewords by a magnitude proportional to the LDPC code girth (\S\ref{s:primer-ldpc:decoder}). \systemacronyms{} distance function distinguishes the true solution of the problem among all the valid LDPC codewords by separating them by a magnitude depending on the distance between the individual codeword and the received information (with channel noise).

\parahead{System Model.} Let ${\textbf{\emph{y}}}$ = [$\emph{y}_\emph{0}$, $\emph{y}_\emph{1}$, ... , $\emph{y}_\emph{N-1}$] be the received information corresponding to an 
LDPC\hyp{}encoded transmitted message $\textbf{\emph{x}}$ = [$\emph{x}_\emph{0}$, 
$\emph{x}_\emph{1}$, ... , $\emph{x}_\emph{N-1}$]. Let \emph{V} be the set of all check constraints $c_{i}$ of this LDPC encoding. Furthermore, let the final decoded message be the final states of the qubits [$\emph{q}_{\emph{0}}$, $\emph{q}_{\emph{1}}$, ... , $\emph{q}_{\emph{N-1}}$] respectively, and let any $\emph{q}_{e_{i}}$ $\forall$ i $>$ 0 be an \emph{ancillary qubit} used for calculation purposes.  Any given binary string is said to be a \emph{valid codeword} when it checks against a given parity check matrix, and an \emph{invalid codeword} otherwise.

\subsubsection{\systemacronyms{} objective function}
\label{s:obj_fcn}

\systemacronyms{} QUBO objective function comprises two terms, an \emph{LDPC satisfier} function $\sum_{\forall c_{i} \in V} L_{sat}(c_{i})$ to prioritize solutions that satisfy the LDPC check constraints (\emph{i.e.}, $L_{sat}(c_{i})$ = 0), and a \emph{distance} function $\sum_{j=0}^{N-1}\Delta_{j}$ to calculate candidate solutions' proximity to the received information. The entire QUBO function is a weighted linear combination of these two terms:
\begin{equation}
\displaystyle\min_{q_{i}}\Big\{W_{1}\displaystyle\sum_{\forall c_{i} \in V} L_{sat}(c_{i}) + W_{2}\displaystyle\sum_{j=0}^{N-1}\Delta_{j}\Big\}
\label{eq:final}
\end{equation}

Here, $W_{1}$ is a positive weight used to enforce LDPC-satisfying constraints, while the positive weight $W_{2}$ increases the success probability of finding the ground truth \cite{ishikawa2009higher}. 

\begin{figure}
\centering
\includegraphics[width=\linewidth]{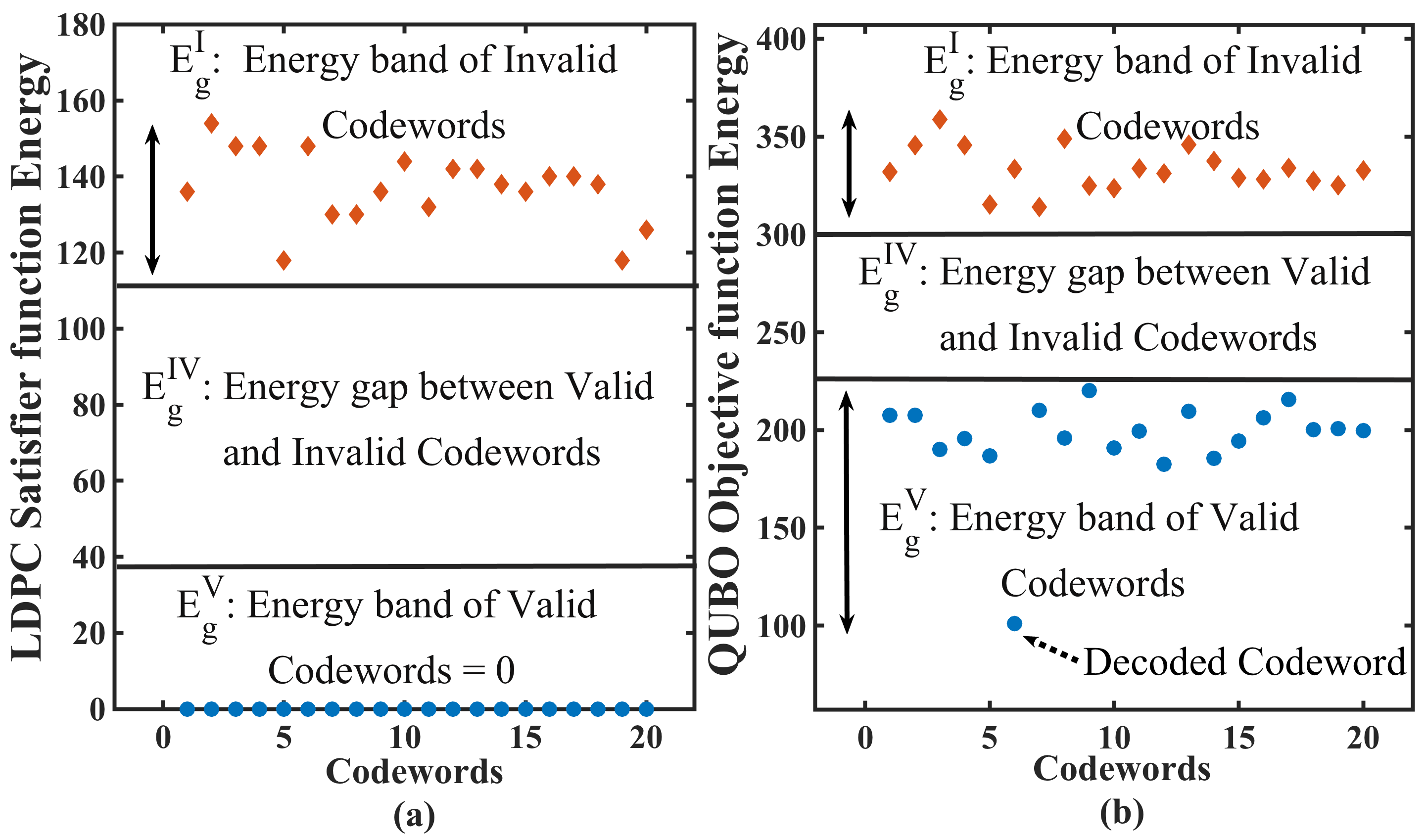}
\caption{{\bf (a)} LDPC satisfier function creating an energy gap between valid and invalid codewords. {\bf (b)} 
\systemacronyms{} objective function seperating
the energy bands of both the valid and invalid LDPC codewords, to correctly decode.}
\label{fig:energy}
\end{figure}

The overall mechanism is depicted in Fig.~\ref{fig:energy} with real data:\ computing the energy values of 20~valid and 20~invalid codewords drawn at random. In Fig. \ref{fig:energy}(a), we see an \emph{energy gap} (whose magnitude is denoted $E_{g}^{IV}$) that our LDPC satisfier function creates between valid and invalid codewords. Note that $E_{g}^{IV}$ is directly proportional to the girth (\S\ref{s:primer-ldpc:decoder}) of the LDPC code (\textit{i.e.}, if the girth of the code is low, there exists an invalid codeword which fails lesser number of check constraints, thus implying a low energy gap $E_{g}^{IV}$).
Increasing $W_{1}$ in Eq.~\ref{eq:final} increases this energy gap, thus eliminating invalid codewords as potential solutions. We observe in Fig.~\ref{fig:energy}(b) that
the distance function distinguishes the actually\hyp{}transmitted codeword 
from other valid (but not transmitted) codewords that would otherwise also land in the ground energy state. The distance function works
by separating the energy levels of both the valid $E_{g}^{V}$ and 
the invalid $E_{g}^{I}$ codewords by a factor proportional to 
the design parameter $W_{2}$. We explore experimentally in \S\ref{sec:eval} the impact of wireless channel SNR and the QA dynamic range on the best choice of $W_1$ and $W_2$.

\subsubsection{LDPC satisfier function} 
\label{s:sat_fcn}

The only LDPC encoding constraint is that the modulo\hyp{}two
bit\hyp{}sum at every check node is zero, \emph{i.e.}, that the sum be even.
For each check node $c_{i}$ we define the function:
\begin{equation}
\textstyle{L_{sat}(c_{i}) = \left(\left(\sum_{\forall j: h_{ij}=1}q_{j}\right) - 
    2L_{e}(c_{i})\right)^2 \forall c_{i} \in V},
\label{eq:F}
\end{equation}
The LDPC constraint is satisfied at check node $c_{i}$, if and only if $L_{sat}(c_{i}) = 0$. Here $L_{e}(c_{i})$ is a function of ancillary qubits $\{q_{e_i}\}$
(defined in \S\ref{s:design:qubo}).
We formulate $L_e$ to use minimal number of ancillary qubits
with the following minimization:
\begin{align}
L_{e}(c_{i}) &= \sum_{s=1}^{t} (2^{s-1}\cdot q_{e_{s+k}})\label{eq:Fe}\\
    t &= \min_{n \in \mathbb{Z}} \{2^{n+1} - 2 \geq d(c_i) - (d(c_i)\mod 2)\}.\label{eq:Fe2}
\end{align}
\vspace{-10pt}
\begin{table}[ht]
\caption{Ancillary qubits required versus check node degree.}
\label{t:exqu}
\begin{small}
\begin{tabularx}{\linewidth}{X*{5}{c}}\toprule
\textbf{Check node degree $\mathbf{d(c_{i})}$:} & 3 & 4--7 & 8--15 & 16--31\\
\textbf{Ancillary qubits required:}& 1 & 2 & 3 & 4\\\bottomrule
\end{tabularx}
\end{small}
\end{table}

where $d(c_{i})$ is the degree of $c_{i}$ (\textit{i.e.}, the number of bits in check constraint $c_i$). In Eq.~\ref{eq:Fe}, the value of $k$ in $L_{e}(c_{i})$ 
is the largest index of the ancillary qubit used while 
computing $L_{e}(c_{i-1})$, ensuring each ancillary qubit is only used once. This formulation of $L_e$ is the binary encoding of integers in the range [0, $2^t-1$], where a single integer corresponds to a single ancillary qubit configuration. The number of ancillary qubits required per check node is given in Table~\ref{t:exqu}.
Upon expansion of Eq.~\ref{eq:F}, $L_{sat}(c_i)$ introduces both biases and couplers to the objective QUBO and hence require embedding on the Chimera graph.

\subsubsection{Distance function}
\label{s:dist_fcn}

We define a distance
$\Delta_{i}$ that computes the proximity of the qubit $q_i$ to its 
respective received information $y_i$ as:
\begin{equation}
\Delta_{i} = \left(q_{i} - Pr(q_{i} = 1|y_i)\right)^2.
\label{eq:delta_mod}
\end{equation}
% \cite{yazdani2011efficient}

In Eq.\ref{eq:delta_mod}, the probability that $q_i$ should take a one value given the received soft information $y_i$, can be computed using the likelihood information obtained from the soft demapping of received symbols, for various modulations and channels \cite{yazdani2011efficient}. For instance, for a BPSK\hyp{}modulated ($0\to -1$, $1 \to +1$) information transmitted over an 
AWGN channel with noise variance $\sigma^2$, this probability is given by 
$1/(1 + e^{-2y_{i}/\sigma^2})$.

Hence, we observe that $\Delta_{i}$ is lesser for the $q_{i} \in$ \{0, 1\} that has a greater probability of being the transmitted bit.
Upon expansion of Eq.~\ref{eq:delta_mod}, we note that the distance function introduces only biases to the QUBO problem and hence do not require embedding due to the absence of coupler terms.

\subsection{Embedding on Annealer Hardware}
\label{sec:5.2}\label{s:design:embedding}

Section~\ref{sec:4} above has described the process of embedding problems onto the QA in general terms.  
In this section, we explain how we embed \systemacronyms{} QUBO reduction onto the Chimera graph of the DW2Q QA hardware. QBP's embedding design can make use of an arbitrarily\hyp{}large hardware qubit connectivity, supporting LDPC code block lengths up to 420 bits on state-of-the-art DW2Q QA.

Let us assign 2D coordinate values $U(x,y)$ to each unit cell in the Chimera graph with the bottom\hyp{}left most unit cell as the origin $U(0,0)$. Here we define terminology:
\begin{itemize}
    \item A Chimera unit cell $U(a,b)$ is said to be a \emph{neighbor} of $U(x,y)$ if and only if $|x-a|+|y-b| = 1$, and let \emph{$\Lambda$(x, y)} denote the set of all neighbors of $U(x,y)$.
    
    \item An \emph{intra\hyp{}cell} embedding is an embedding where both participating qubits lie in the same Chimera unit cell.
    
    \item An \emph{inter\hyp{}cell} embedding is an embedding where one of the qubits belongs to $U(x,y)$, and the other participating qubit belong to a unit cell in \emph{$\Lambda$(x, y)}.

\end{itemize}

\paragraph{QGEM: \systemacronyms{} Graph Embedding Model.} We structure our embedding scheme into two levels, \textit{Level I} (\S\ref{s:design:embedding:1})
and \textit{Level II} (\S\ref{s:design:embedding:2}). \systemacronyms{} graph embedding model \textit{(\sysembedder{})} first maps the check constraints (\textit{i.e.}, $L_{sat}$($c_i$)) by constructing the Level-I embedding for all the available Chimera unit cells, and next it accommodates more check constraints via the Level-II embedding, using the idle qubits that were left out during the Level-I embedding. \sysembedder{} makes use of the entire qubit computational resources available in the DW2Q QA hardware leaving no qubit idle in the machine.

\begin{figure}
\centering
\includegraphics[width=\linewidth]{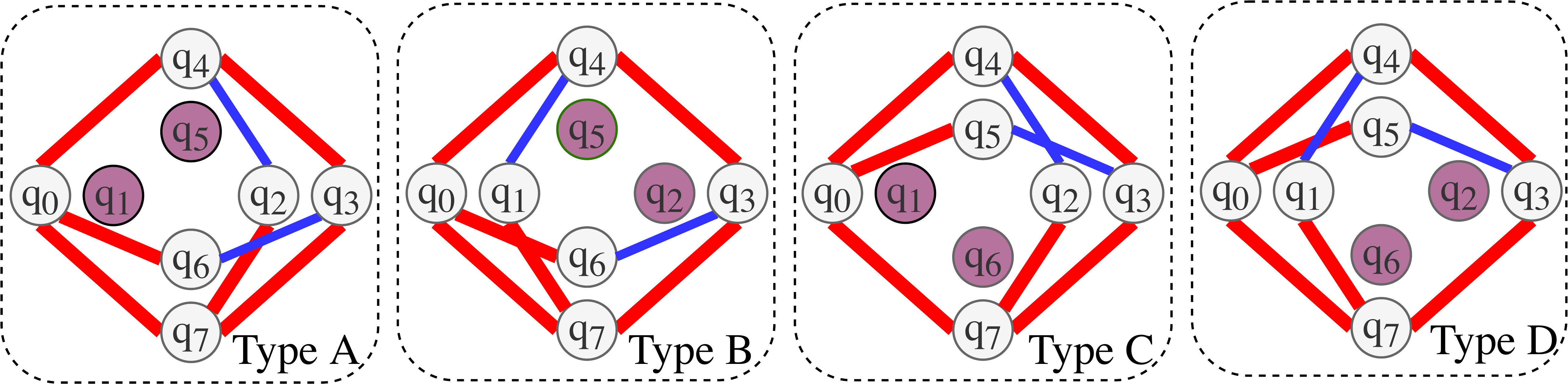}
\caption{\systemacronyms{} unit cell schemas for \emph{Level-I} Chimera Graph embedding. Here ($q_{a}, q_{b}, q_{c}, q_{e_{1}}$) of Eq.~\ref{eq:example} can be interpreted as ($q_{0}, q_{4}, q_{7}, q_{3}$) respectively in each schema. Idle qubits are shown in a darker shade. Embeddings are \textit{thin-blue} lines and \textit{thick-orange} lines are QUBO problem couplers.}
\label{fig:types}
\end{figure}

In the Level-I embedding, \sysembedder{} represents a single check constraint (\textit{i.e.}, each $L_{sat}$($c_i$)) of at most degree three on a single Chimera unit cell using one of the four schemas presented in Fig.~\ref{fig:types}, which we refer to as \textit{Types A--D}. Each of these schemas uses six qubits for a degree\hyp{}three check constraint, leaving two qubits in the unit cell idle. Based on the coordinate location \textit{U(x,y)} of the unit cell, QGEM chooses a single schema for a single Chimera unit cell in a fashion that creates a specific pattern of idle qubits in the Chimera graph, then leverages this pattern to accommodate more check constraints as explained in \S\ref{s:design:embedding:2}. Next, \sysembedder{} places the check constraints that share a common bit closest to each other, then embeds the qubits representing this shared bit to make them agree, as described in Fig.~\ref{fig:embex} (\S\ref{s:qa-primer:logical}). Specifically, if a check constraint $c_i$ is placed in $U(x_0, y_0)$, then \sysembedder{} places the check constraints that share common bits with $c_i$ in $\Lambda(x_0, y_0)$ and embeds the qubits representing such commonly shared bits via an inter-cell embedding (see dotted lines in Fig. \ref{fig:l1}(a)). 

In the Level-II embedding, \sysembedder{} represents a single check constraint in an \textit{ensemble} of nine Chimera unit cells using the pattern of idle qubits that the Level-I embedding leaves. The placement of each of these ensembles in the Chimera graph follows a similar fashion as in Level-I embedding (\textit{i.e.}, placing the ensembles whose Level-II check constraints share bits close to each other).

We detail the overall working of \systemacronyms{} graph embedding model more fully with a running example. Consider a $(2,3)$\hyp{}regular LDPC code: as the degree of each check node is three, let us assume that [$x_{a}, x_{b}, x_{c}$] are the three bits participating in one of the check constraints $c_{i}$. Let [$q_{a}, q_{b}, q_{c}$] be the \textit{bit\hyp{}node\hyp{}representing} qubits used at the decoder to extract [$x_{a}, x_{b}, x_{c}$] respectively. From Eqs. \ref{eq:F} and~\ref{eq:Fe}, the LDPC satisfying constraint of this check node is:
\begin{equation}
L_{sat}(c_{i}) = \big(q_{a} + q_{b} + q_{c} - 2q_{e_{1}}\big)^2
\label{eq:example}
\end{equation}

\subsubsection{Level-I Embedding}
\label{s:design:embedding:1}

\begin{figure}
\centering
\includegraphics[width=\linewidth]{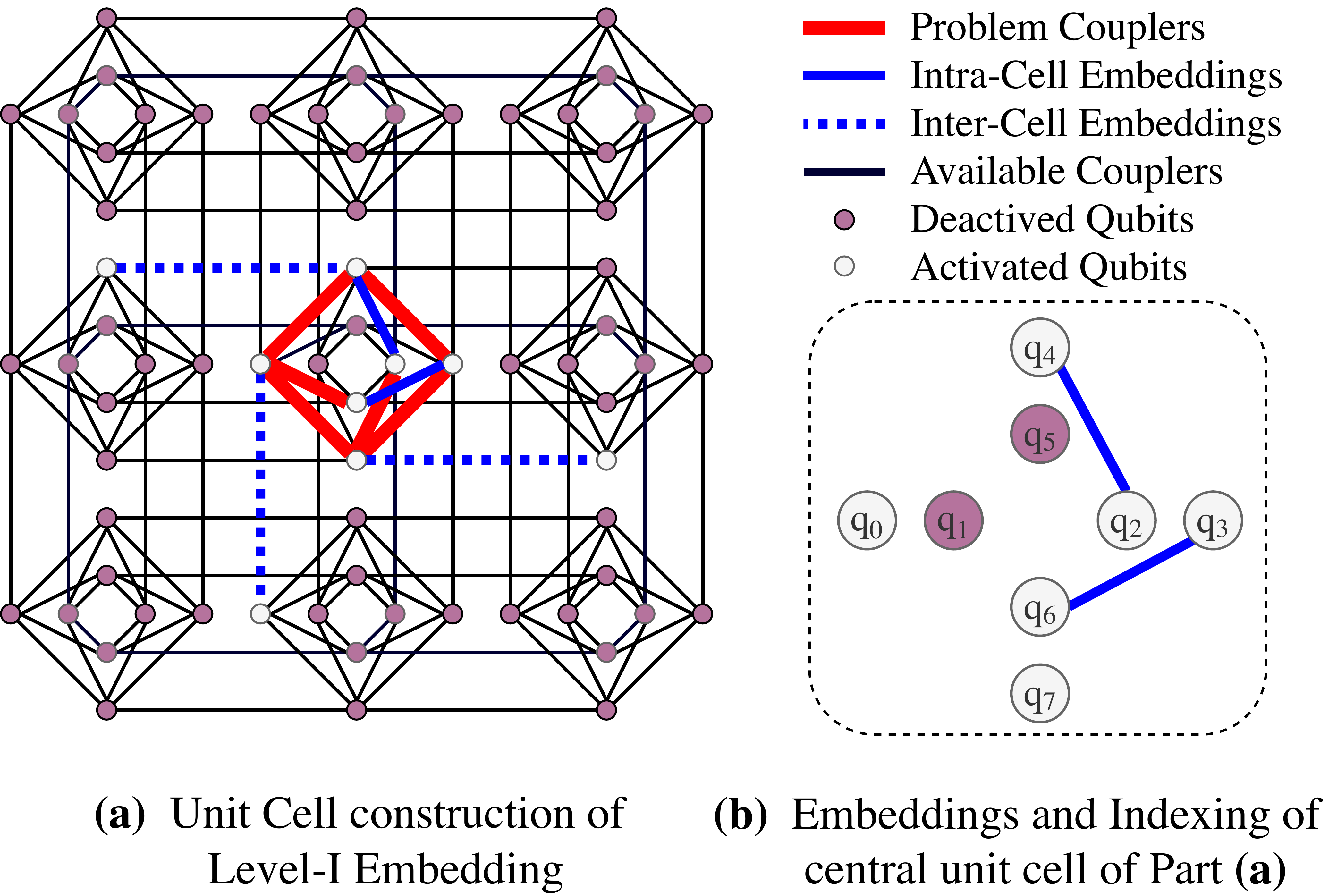}
\caption{QBP's Level-I Chimera Graph Embedding.}
\label{fig:l1}
\end{figure}

Upon expansion of Eq.~\ref{eq:example}, we observe that the quadratic terms (\textit{i.e.}, qubit-pairs) requiring a coupler connectivity are \{~($q_{a}$, $q_{b}$), ($q_{a}$, $q_{c}$), ($q_{a}$, $q_{e_{1}}$), ($q_{b}$, $q_{c}$), ($q_{b}$, $q_{e_{1}}$), ($q_{c}$, $q_{e_{1}}$)~\}. \systemacronyms{} Level-I embedding for the example in Eq. \ref{eq:example} can be visualized by interpreting ($q_{a}$, $q_{b}$, $q_{c}$, $q_{e_{1}}$) as equivalent to ($q_{0}$, $q_{4}$, $q_{7}$, $q_{3}$) respectively in Figs. \ref{fig:types} and \ref{fig:l1}. \systemacronym{} realizes the above required coupler connectivity in four schemas presented in Fig. \ref{fig:types}. We next demonstrate the Type A schema.

\parahead{Construction.} We construct the required-and-available coupler connectivity using the QA's direct physical couplers (\textit{e.g.}, $q_{0}$ to $q_{4}$ in Type A, Fig.~\ref{fig:types}), and realize the required-but-unavailable coupler connectivity \big\{($q_{0}, q_{3}$), ($q_{4}$, $q_{7}$)\big\}, using two intra\hyp{}cell embeddings (\emph{e.g.} $q_2$ to $q_4$ in Type A, Fig. \ref{fig:types}).

\parahead{Placement.} Let us assume that \sysembedder{} chooses the above Type~A schema for one of the Chimera unit cells whose placement is shown in Fig.~\ref{fig:l1}(a).~We note that the example LDPC code is (2, 3)-regular, and so every bit node participates in two check constraints.~This implies that each bit-node-representing qubit (\textit{i.e.}, excluding ancillary qubits) must be present in two Chimera unit cells since in the Level\hyp{}I embedding, we represent a check constraint in a single Chimera unit cell. \sysembedder{} thus represents the other check constraint of each of these bit\hyp{}node\hyp{}representing qubits \big\{$q_{0}$, $q_{4}$, $q_{7}$\big\} in a neighbor unit cell connected via an inter\hyp{}cell embedding as depicted in Fig.~\ref{fig:l1}(a), thus making the physical qubits involved in the embedding agree. \sysembedder{} repeats this construction over the entire Chimera graph, mapping each check constraint to an appropriate physical location in the QA hardware. \sysembedder{} selects the schema type to use (see Fig.~\ref{fig:types}) for each unit cell in a way that the two idle qubits of the Level-I unit cell schemas form the pattern as shown in Fig.~\ref{fig:l2}(a).

\subsubsection{Level-II Embedding}
\label{s:design:embedding:2}

Let us continue with the example of Eq.~\ref{eq:example}. The overview of \systemacronyms{} Level-II embedding is presented in Fig.~\ref{fig:l2}. Here, in Level-II, the mapping of bits in the check constraint of Eq.~\ref{eq:example} to physical qubits is ($q_{a}$, $q_{b}$, $q_{c}$, $q_{e_{1}}$) map to ($q_{A}$, $q_{B}$, $q_{C}$, $q_{E}$) respectively. In the Fig.~\ref{fig:l2}, qubits $q_{A_{i}}$ $\forall$ i $\in$ [0, 3] represent $q_{A}$, $q_{B_{i}}$ $\forall$ i $\in$ [0, 2] represent $q_{B}$, $q_{C_{i}}$ $\forall$ i $\in$ [0, 2] represent $q_{C}$, and the qubits $q_{E_{i}}$ $\forall$ i $\in$ [0, 3] represent $q_{E}$, as they are embedded together as shown in Fig.~\ref{fig:l2}(b). The pattern in the figure now allows us to realize all the required coupler connectivity of the example in Eq.~\ref{eq:example} as depicted in Fig.~\ref{fig:l2}(a). Similar to our Level-I placement policy, QGEM repeats this construction over the entire Chimera graph, mapping each Level-II check constraint to an appropriate physical location in the QA.

%% file: implementation.tex
\section{Implementation}
\label{sec:impl}

We implement \systemacronym{} on the DW2Q QA: our decoder targets a $(2, 3)$\hyp{}regular maximum girth LDPC code of block length 420~bits. In the DW2Q, a \textit{solver} is a resource that runs the input problem.  We implement \systemacronym{} remotely via the \textit{C16-VFYC} hardware solver API, using the Python client library. This solver first maps the implementation of the problem at hand directly onto the DW2Q's QPU hardware, then determines the final states of the few (15 on our particular DW2Q) defective qubits via post\hyp{}processing on integrated conventional silicon \cite{dwavedocs}. Since post\hyp{}processing problem size is two orders of magnitude smaller than overall problem size, post\hyp{}processing parallelizes with annealer computation and therefore does not factor into overall performance. 

DW2Q readout fidelity is greater than 99\%, and the chance of QPU programming error is less than 10\% for problems that use all the available QA hardware resources \cite{dwavetech}. However, we increase readout fidelity and decrease the chance of programming error via the standard method of running multiple anneals for every LDPC decoding problem, where each anneal reads the solution bit-string once. In our evaluation, we further quantify the unavoidable \textit{intrinsic control errors} (\S\ref{s:discussion}) that arise due to the quantization effects and the flux noise of the qubits \cite{dwavetech}.   Our end\hyp{}to\hyp{}end evaluation results capture all the above sources of QA imprecision.

\begin{figure}
\centering
\includegraphics[width=\linewidth]{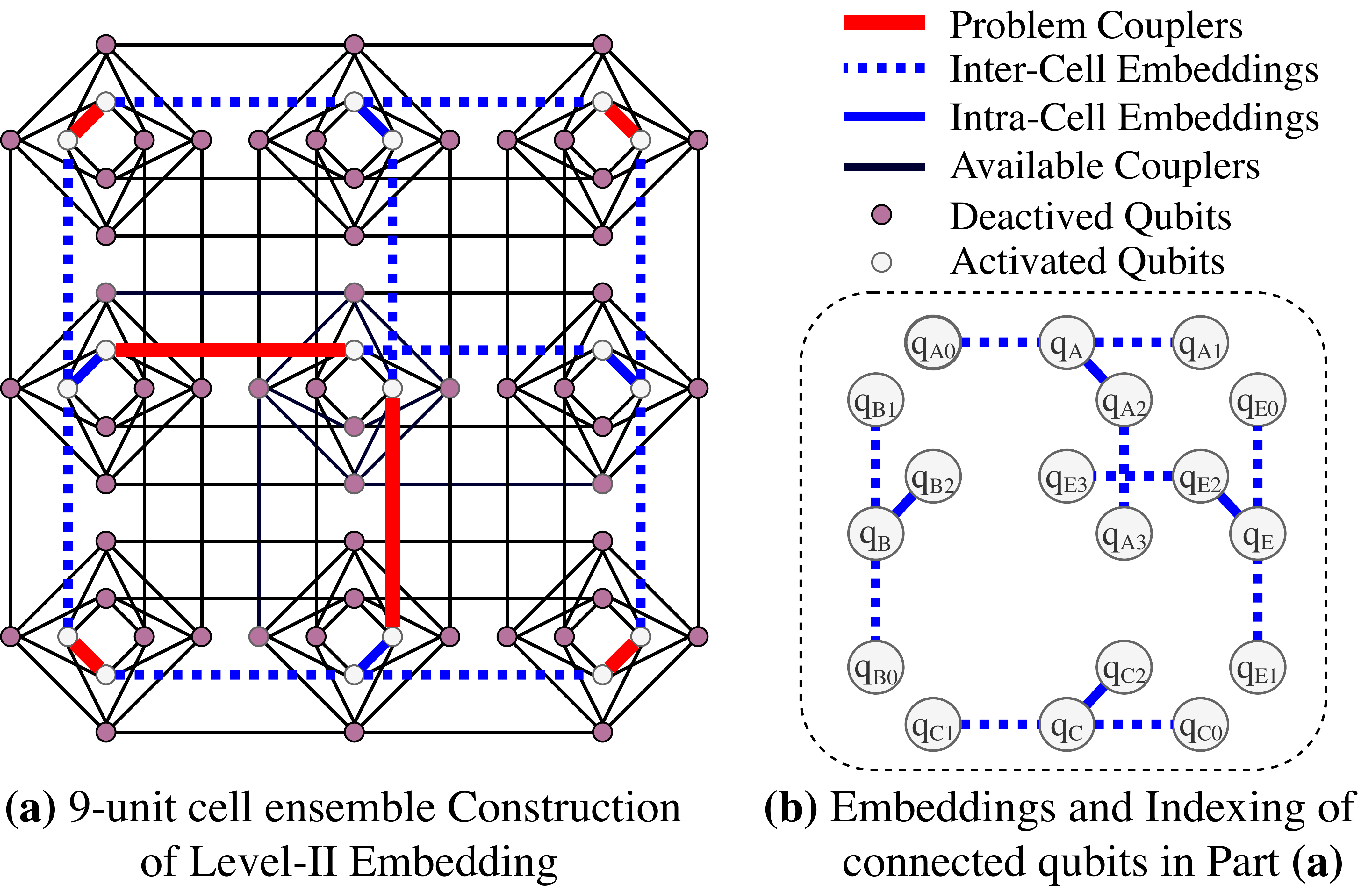}
\caption{QBP's Level-II Chimera Graph Embedding. In Fig.~\ref{fig:l2}(b), ($q_{A}, q_{B}, q_{C}, q_{E}$) represent ($q_{a}, q_{b}, q_{c}, q_{e_{1}}$) of Eq.~\ref{eq:example}.}
\label{fig:l2}
\end{figure}

%% file: evaluation.tex
\section{Evaluation}
\label{sec:7}\label{sec:eval}

Our experimental evaluation is on the DW2Q QA, beginning with our experimental methodology description (\S\ref{s:eval:methodology}).  We measure performance over a variety of DW2Q parameter settings (chosen in \S\ref{s:eval:micro}), and in both simulated wireless channels, and realistic trace-driven wireless channels.  End\hyp{}to\hyp{}end experiments (\S\ref{s:eval:system}) compare head\hyp{}to\hyp{}head against FPGA-based soft belief propagation decoding.

\subsection{Experimental Methodology} 
\label{s:eval:methodology}

Let us define an \textit{instance I} as an LDPC codeword. Our evaluation dataset consists 150 instances with an overall $2\times 10^4$ message bits. We conduct $10^4$ anneals for each instance and note the distribution of the solutions returned along with their occurrence frequency. If $N_{s}^{I}$ is the number of different solutions returned for an instance \textit{I}, we rank these solutions in increasing order of their energies as $R_{1}, ... , R_{N_{s}^{I}}$ with $R_{1}$ being the rank of the minimum energy solution. All the $N_{s}^{I}$ solutions can be treated as identically independent random variables, as each anneal is identical and independent.

\subsubsection{BER Evaluation} Let $R_{min}$ be the rank of the minimum energy solution in a particular population sample of the entire solution distribution, of size $N_a$ (<$10^4$) anneals. We compute the expected \textit{number of bit errors} $N_B^I$ of an instance \textit{I} over $N_a$ anneals as:
\begin{equation}
E\big[N_B^I\big| N_{a}] = \displaystyle\sum_{i=1}^{N_s^I}\Pr\big(R_{min}=R_{i} | I, N_a\big)\cdot N_{B}^{I}\big(R_{i} | I, N_a\big),
\label{eq:Eber}
\end{equation}

\noindent
where the probability of $R_{min}$ being $R_{i}$ $\forall$ i $\in$ [1, $N_{s}^{I}$] for an instance \textit{I}, over performing $N_{a}$ anneals is computed using the cumulative distribution function $F(\cdot)$ of observed solutions in $10^4$ anneals as \cite{kingman1975random}:
\setlength{\abovedisplayskip}{5pt}
\setlength{\belowdisplayskip}{5pt}
\begin{equation}
\Pr\big(R_{min} = R_{i}| I, N_{a}\big) = \big(1 - F(R_{i-1})\big)^{N_{a}} - \big(1 - F(R_{i})\big)^{N_{a}},
\label{eq:probRmin}
\end{equation}

\noindent
Hence we compute the bit error rate (BER) of an instance \textit{I} with K information bits upon performing $N_{a}$ anneals as:
\setlength{\abovedisplayskip}{5pt}
\setlength{\belowdisplayskip}{5pt}
\begin{equation}
\mbox{BER} = E\big[N_B^I\big| N_{a}]/K.
\label{eq:ber}
\end{equation}

\subsubsection{FER Evaluation}

\textit{Frame Construction:} We construct a frame of length $N_F$ using data blocks of length $N_B$, so we require $N_F/N_B$ such blocks, where each block is an instance. If $N_{ins}$ is the number of available instances, we can construct a single frame by combining any $N_F/N_B$ instances among the available $N_{ins}$ instances. Thus the total number of distinct frames we construct for our \textit{frame error rate} (FER) evaluation is $\binom{N_{ins}}{N_F/N_B}$.

\noindent
\textit{FER Calculation:} A frame is error-free iff all the code blocks in the frame has zero errors, just as if it has a cyclic redundancy check appended. We compute the probability of a particular $k^{th}$ frame being error-free (~$Pr(F_{ef}^k)$~) as:
\begin{equation}
\Pr\left(F_{ef}^k\right) = \prod_{I=1}^{N_F/N_B}\left\{\sum_{\forall i} \Pr\left(R_{min} = R_{i}| I, N_a, N_B^I(R_{i})=0\right)\right\}
\label{eq:fer_na}
\end{equation}

Then we compute the overall frame error rate (FER) as:
\setlength{\abovedisplayskip}{5pt}
\setlength{\belowdisplayskip}{5pt}
\begin{equation}
\mbox{FER} = \left[\sum_{k=1}^{\binom{N_{ins}}{N_F/N_B}}\left\{1 - \Pr(~F_{ef}^k~)\right\}\right]\big/\binom{N_{ins}}{N_F/N_B}
\label{eq:fer}
\end{equation}

\subsubsection{Wireless Trace-driven Evaluation}

We collected channel traces from an aerial drone server communicating with a ground client in an outdoor environment, using the Intel 5300 NIC wireless chip at the client \cite{halperin2011tool}. In realistic wireless transmissions, code blocks are transmitted over multiple OFDM symbols, where subcarriers within an OFDM symbol typically experience a diverse range of channels. In our performance evaluation over experimental channels, we compute the per-subcarrier SNR information through \textit{channel state information} (CSI) readings, and distribute a corresponding Gaussian noise over bits individually for every subcarrier. Next we demodulate and interleave the data symbols and perform \systemacronyms{} decoding. Hence we use the distance function (\S5.1.3) for this evaluation with $\sigma^{2}$ equal to the noise variance experienced by $y_i$'s subcarrier.

\subsubsection{QA versus FPGA Throughput Evaluation}
\label{s:fpga_throughput_eval_meth}

Consider a data frame with \textit{$N_K$} message bits. 
Let us assume that QBP decodes this frame on the
QA for a $T_c$ \textit{compute time}, and soft BP decodes the same frame on an 
FPGA with clock frequency $f_{clk}$, for $N_{it}$ iterations. 
Let $N_{clk/it}$ be the number of FPGA clock cycles the 
soft BP requires to complete an iteration. The actual throughput 
QBP achieves is then $(1 - FER_{QA})\cdot N_K/T_c$, and the
actual FPGA soft BP\hyp{}based throughput is then
$(1 - FER_{FPGA})\cdot N_K\cdot f_{clk}/ (N_{it}\cdot N_{clk/it})$.

The values of $N_{clk/it}$ and $f_{clk}$ depend on the decoder implementation 
architecture (\textit{i.e.,} serial or parallel) and FPGA hardware type. 
In order to make a throughput comparison between QA and FPGAs, we 
evaluate the QA throughput versus the best silicon realization 
(\textit{i.e.,} a fully-parallel decoder, $N_{clk/it}$ = 1) 
throughput on the highest specification Xilinx FPGA,
for a range of FPGA clock frequencies and highlight the design\hyp{}dependent 
operating\hyp{}time regions (\S\ref{s:eval:system}).

\subsection{Parameter Sensitivity Analysis}
\label{s:eval:micro}

In this section, we determine DW2Q QA's optimal system parameters, including JFerro ($\abs{J_F}$), annealing time ($T_a$), number of anneals ($N_a$), and the design parameter $W_2$ for evaluating \systemacronyms{} overall end-to-end system performance (\S\ref{s:eval:system}).

\subsubsection{Choice of Embedding Coupler Strength $|J_F|$} In the QA literature, the coupler strength of an embedding is termed JFerro (\S\ref{s:primer-qa}). As the fundamental purpose of embeddings is to make qubits agree, a large, negative JFerro is required in order to ensure the embedding is effective (\S\ref{s:primer-qa:forms}). However, as the supported range for coupler strengths in DW2Q QA is $[-1,1]$, it is general practice to normalize all QUBO coefficients with respect to $\abs{J_F}$ to bring all the coupler strengths into this supported range $[-1,1]$.
\begin{figure}
\centering

    \begin{subfigure}[b]{0.5\linewidth}
		\centering
		\includegraphics[width=\textwidth]{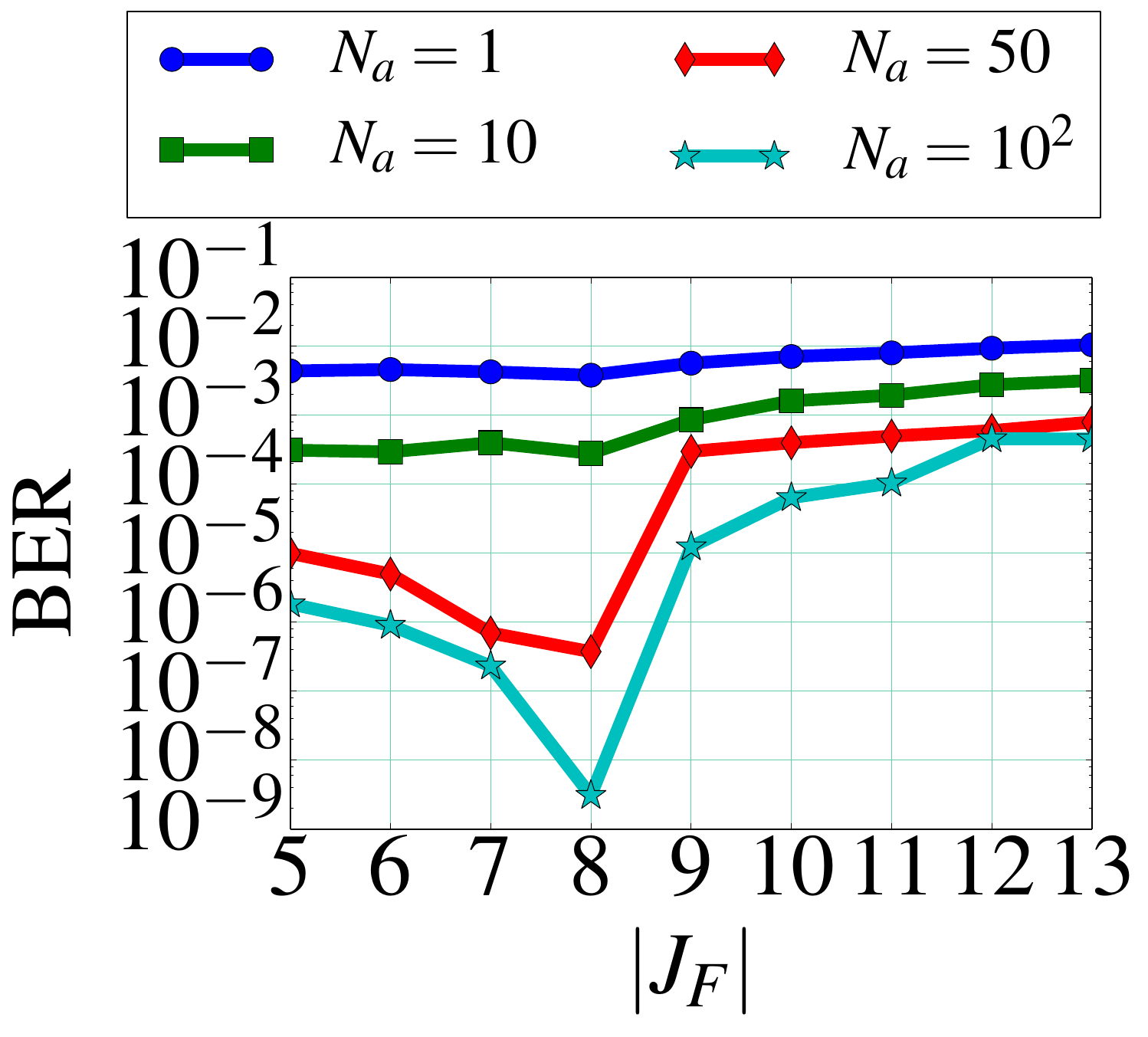}
	\end{subfigure}\hfill
	\begin{subfigure}[b]{0.5\linewidth}
		\centering
		\includegraphics[width=\textwidth]{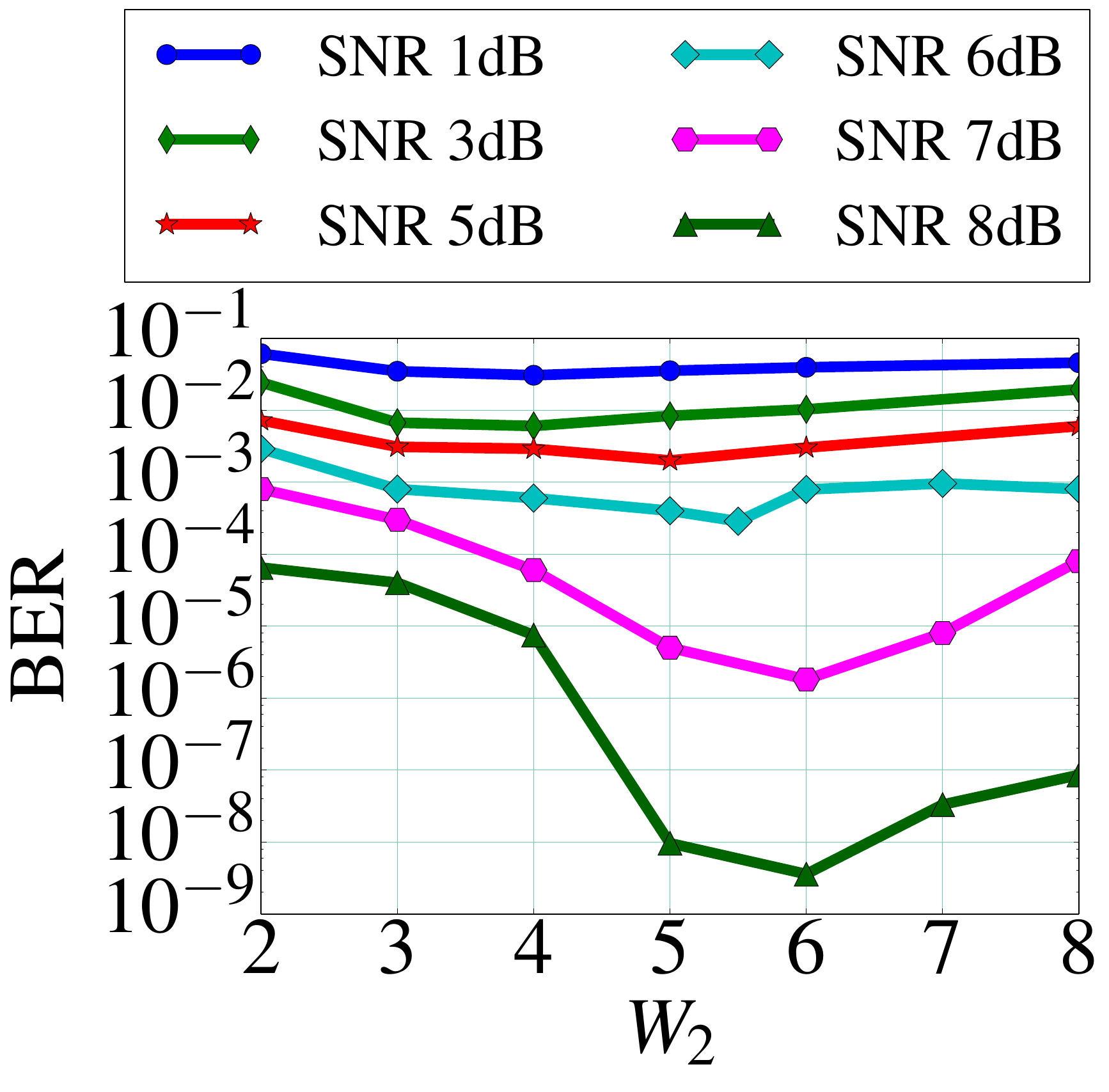}
	\end{subfigure}

\caption{\textit{Left}. Choosing \textit{JFerro} strength to minimize BER. \textit{Right}. Effect of $W_{2}$ on BER at various channel SNRs. The magnitude of $W_2$ that minimizes BER is proportional to SNR.}
\label{fig:JW}

\end{figure}

Consider a QUBO problem with coupler strengths in the range [A, B]. Then $\abs{J_F}$ must be greater than max($\abs{A}$, $\abs{B}$) to prioritize embeddings over problem couplers, and moderate enough to distinguish the new normalized coupler strengths \big[$\frac{A}{\abs{J_F}}$, $\frac{B}{\abs{J_F}}$\big] as the range lessens. We perform our JFerro sensitivity analysis at a moderate SNR of 8~dB. We use a relatively high anneal time ($T_a$ = 299~$\mu$s), to ensure minimal disturbance from the time limit, and we choose our QUBO design parameters $W_1$ = 1.0 and $W_2$ = 6.0, experiments show that all other values of $W_1$ and $W_2$ results in similar trends for the JFerro sensitivity. Fig.~\ref{fig:JW} (\textit{left}) depicts \systemacronyms{} BER performance at various $\abs{J_F}$ strengths. The BER curve of $N_{a}=\{50, 10^2\}$ anneals clearly depict that $\abs{J_F}$ = 8.0 minimizes BER, while for $N_{a}$ = \{1, 10\} anneals BER is barely minimized at $\abs{J_F}$ = 8.0, as the effect of $\abs{J_F}$ is slight because of fewer anneals. Hence heretofore we set $\abs{J_F}$ = 8.0 for further evaluation.

\noindent
\subsubsection{Choice of design parameter $W_2$}

QBP's LDPC satisfier function (Eq.~\ref{eq:F}) introduces coupler strengths (\textit{i.e.}, quadratic coefficients) greater than one, and hence must be normalized to bring all the problem coupler strengths into the supported $[-1, 1]$ range. Hence we set $W_1$ = 1.0 and consider the choice of $W_{2}$, the parameter that determines sensitivity to the received bits, in order to identify the correct codeword. We find the optimal value for $W_2$ dynamically with the wireless channel SNR, to balance between the received soft information and the LDPC code constraints.

We perform our $W_2$ sensitivity analysis at $\abs{J_F}$ = 8.0 (\S7.2.1), $W_1$ = 1.0 (\S7.2.2), and use a high anneal time ($T_a$ = $299$~$\mu$s), to ensure minimal disturbance from the time limit. Fig.~\ref{fig:JW} (\textit{right}) depicts \systemacronyms{} BER performance at various SNRs while varying $W_2$.  In the figure we observe that the magnitude of $W_2$ that minimizes BER, increases with increase in channel SNR. Hence \systemacronym{} chooses $W_{2}$ at the time of data reception. As an incoming frame arrives, the receiver uses the packet preamble to estimate SNR, and then looks up the best $W_{2}$ for decoding in a lookup table.

\begin{figure}
\centering
\includegraphics[width=\linewidth]{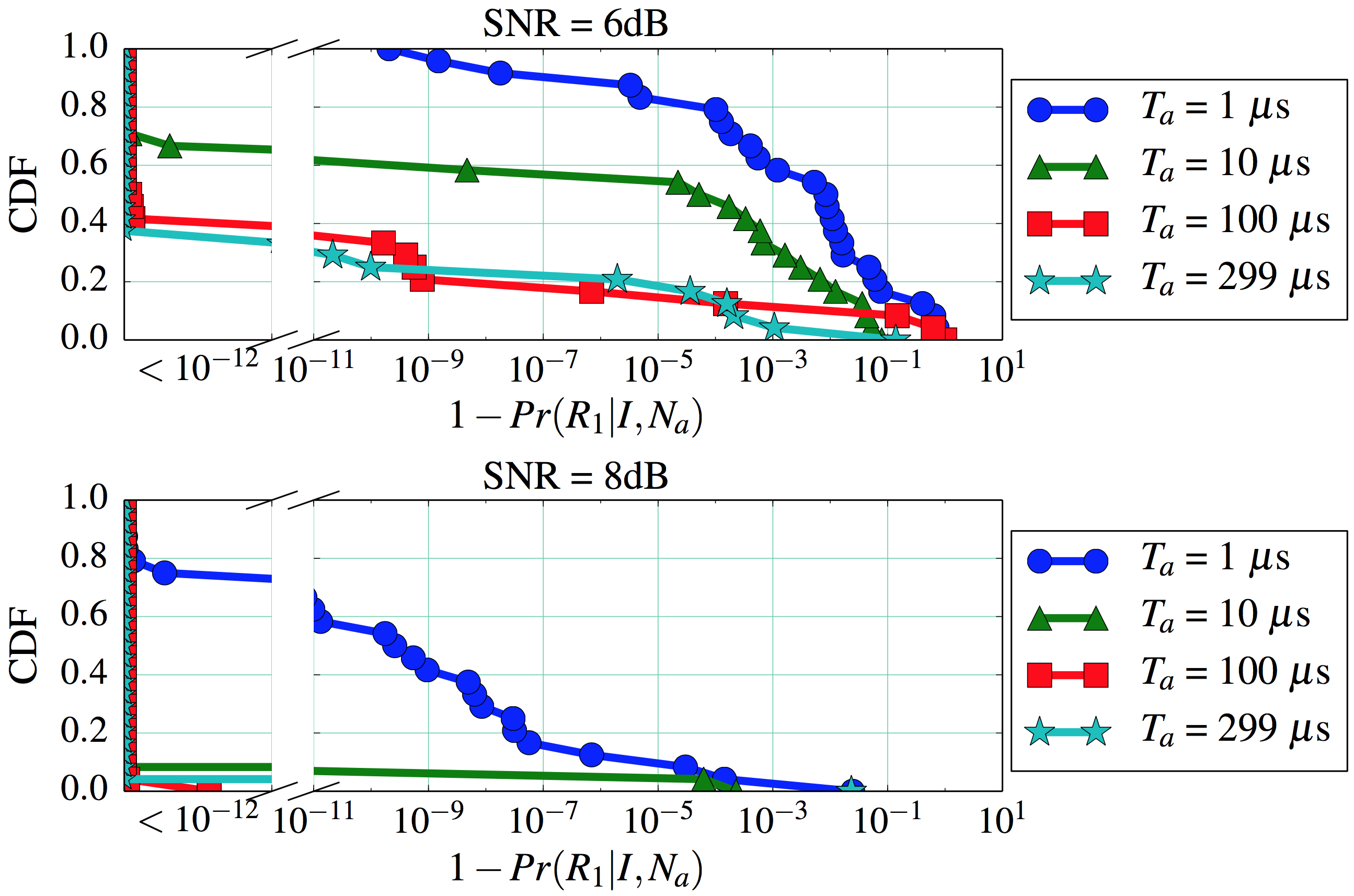}
\caption{Choosing anneal time $T_{a}$. Figure depicts the probability of \textbf{not} finding the ground truth across distribution of problem instances. $T_a$ = 1 $\mu$s is sufficient to achieve a high probability of finding ground truth.}
\label{fig:tan}
\end{figure}

\subsubsection{Choosing the annealing time $T_a$}

We perform our annealing time sensitivity analysis using $\abs{J_F}$ = 8.0 (\S7.2.1) and $W_1$ = 1.0 (\S7.2.2). We choose $W_2$ as above (\S7.2.2) and perform $N_a$ = 10 anneals (any number of anneals results in similar trends).
Fig.~\ref{fig:tan} presents the probability of {\bf not} finding the minimum energy solution over the cumulative distribution across problem instances. We find that an anneal time as low as one~$\mu$s yields a high probability of finding the ground truth, hence we consider $T_a$ = 1~$\mu$s.

\begin{figure}
\centering
\includegraphics[width=\linewidth]{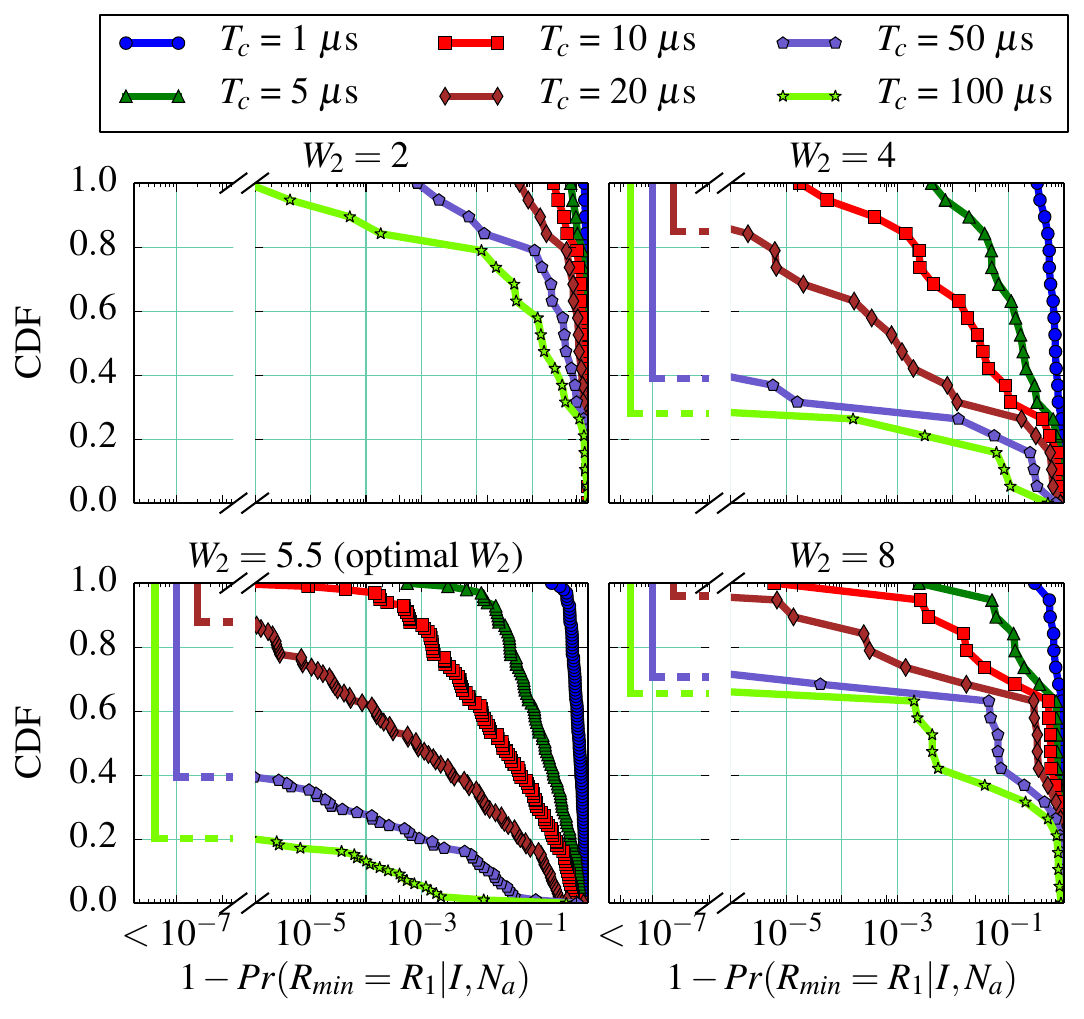}
\caption{The effect of calibrations in (\S7.2) at SNR 6 dB, depicting the probability of \textbf{not} finding the minimum energy state at $|J_F|=8.0$. All plots share common x-y axes, and the distribution is across problem instances. The bottom--left plot corresponds to the best $W_2$ at SNR 6 dB (see Fig. \ref{fig:JW} \textit{right}).}
\label{fig:w2}
\end{figure}

Heretofore we quantify \systemacronyms{} performance over total \textit{compute time} $T_c$, where $T_c$ = $N_a\cdot T_a$. Fig.~\ref{fig:w2} depicts the combined result of the overall calibrations presented in (\S7.2). Specifically, Fig.~\ref{fig:w2} shows the probability of \textbf{not} finding the minimum energy solution across the cumulative distribution of problem instances at wireless channel SNR 6~dB over various choices of $W_2$ and computing times ($T_c$). The figure shows that the best choice of $W_2$ results in a relatively low probability of \textbf{not} finding the ground truth, as well as the benefits of increasing compute time up to 100~$\mu$s.

\begin{figure*}[ht]
\centering
\begin{subfigure}[b]{0.32\linewidth}
\includegraphics[width=\linewidth]{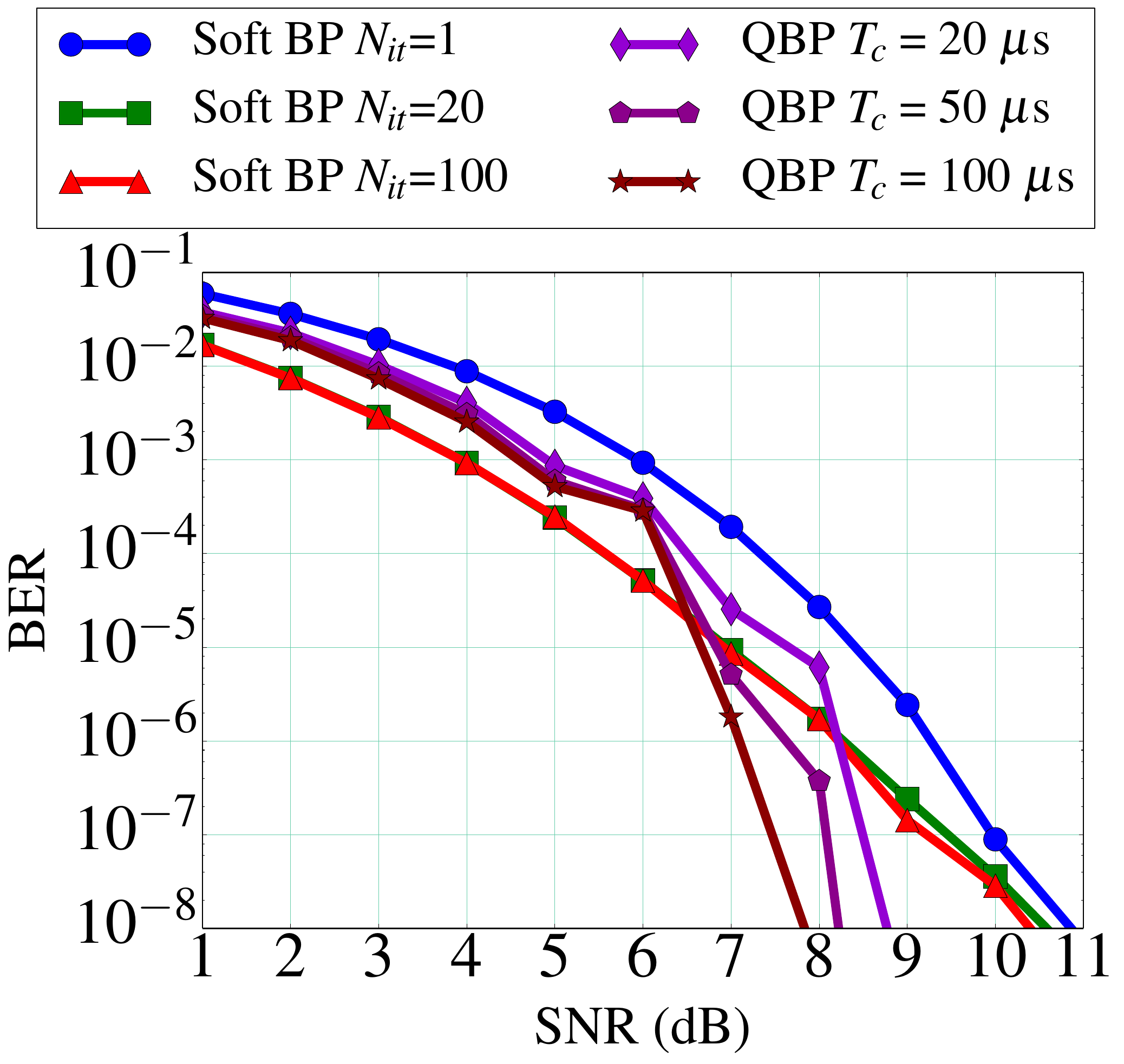}
\caption{Average BER, AWGN channel.}
\label{fig:ber_sim}
\end{subfigure}
\hfill
\begin{subfigure}[b]{0.32\linewidth}
\includegraphics[width=\linewidth]{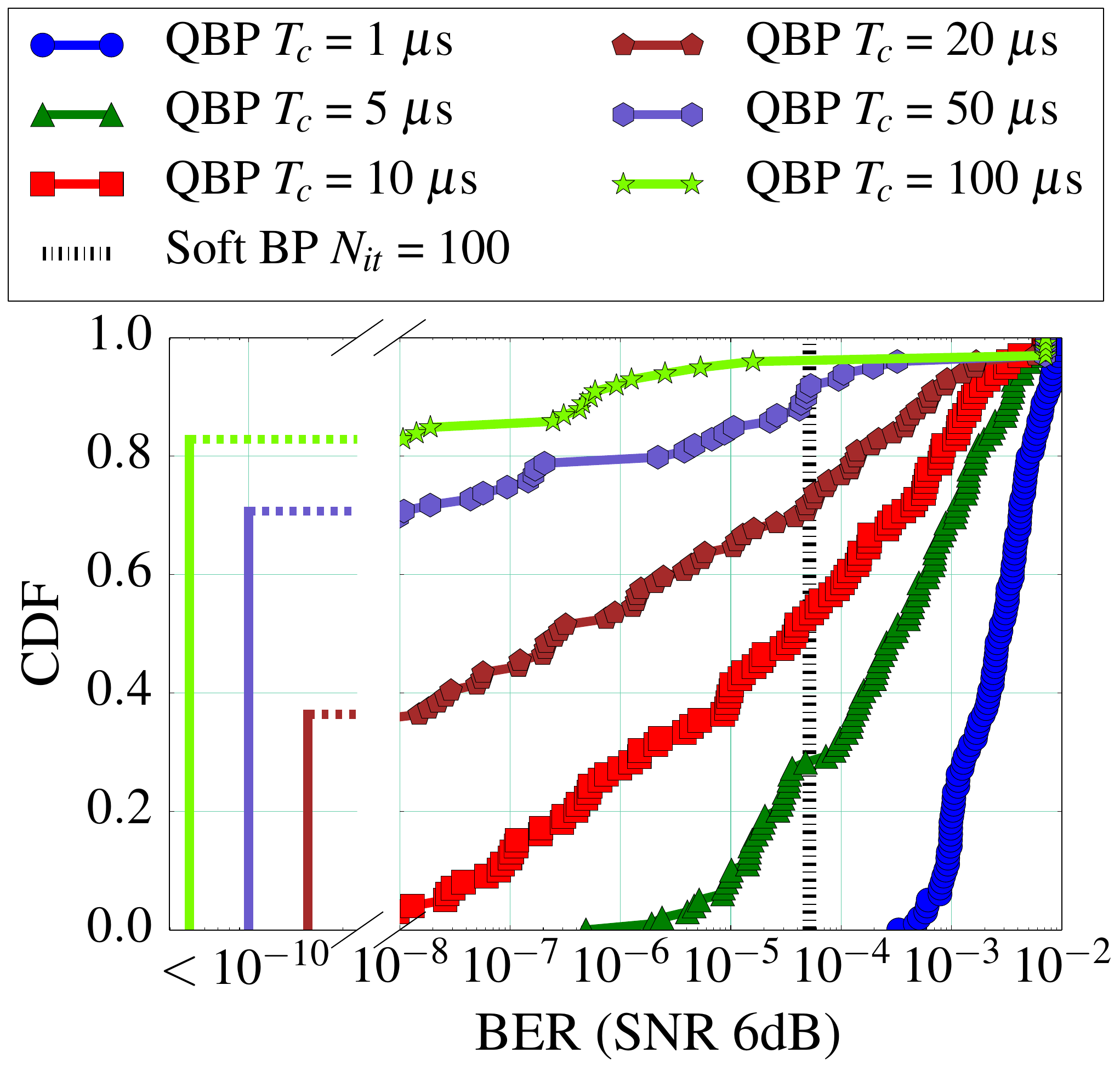}
\caption{CDF, AWGN channel.}
\label{fig:cdf_sim}
\end{subfigure}
\hfill
\begin{subfigure}[b]{0.32\linewidth}
\centering
\includegraphics[width=\linewidth]{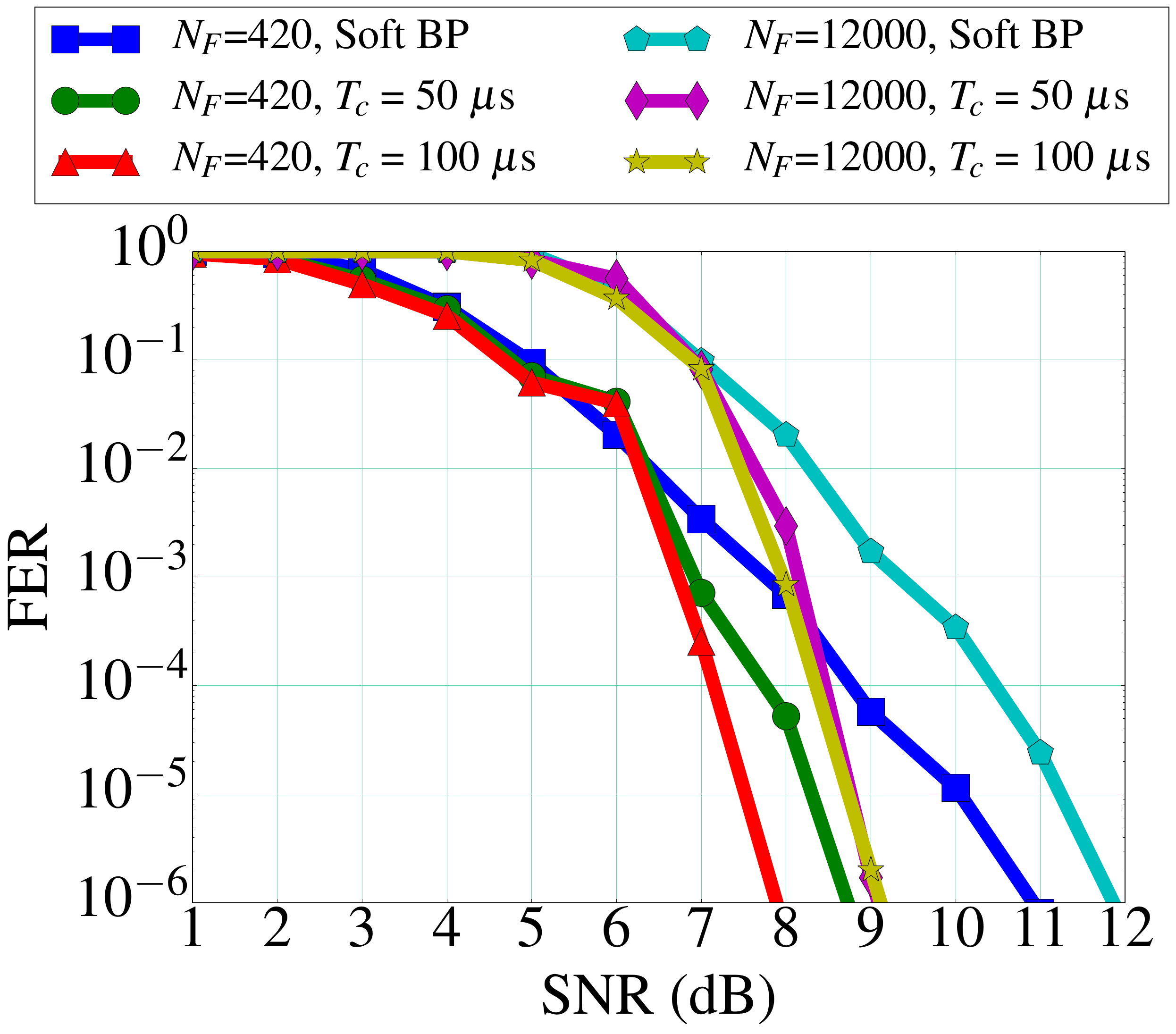}
\caption{Average FER, AWGN channel.}
\label{fig:fer_sim}
\end{subfigure}
\hfill
\vspace{5pt}
\begin{subfigure}[b]{\linewidth}
\includegraphics[width=\linewidth]{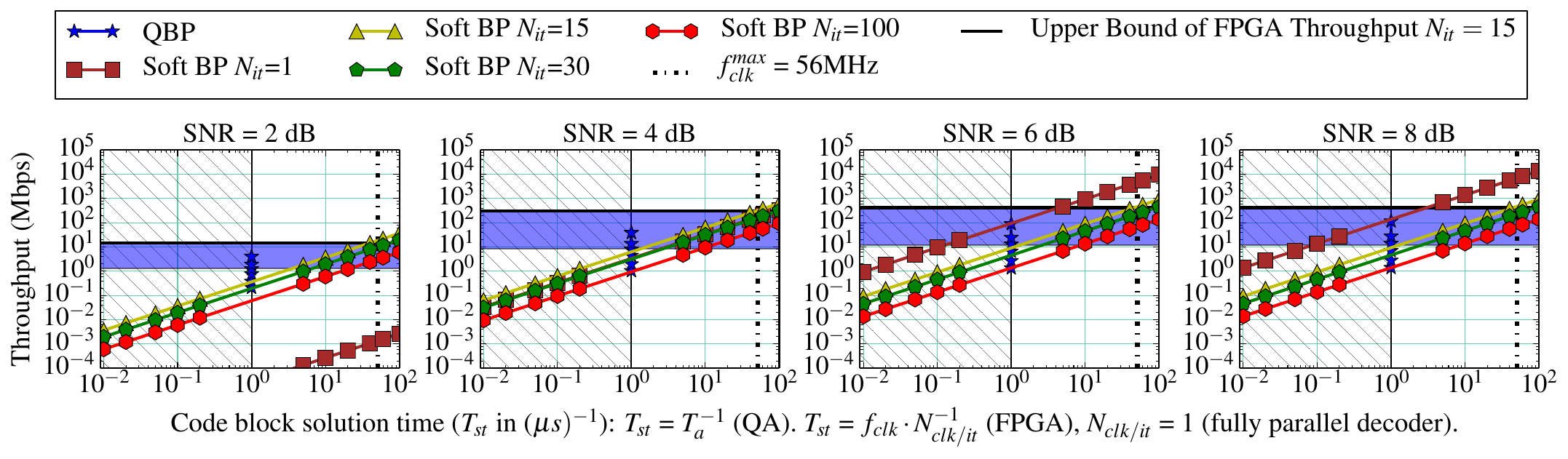}
\caption{Throughput comparison of QBP versus soft BP decoders for a (2,3)-regular code of block length 420 bits. In the figure, the hatched area is the operating-time region of QA, the colored (solid filled) area is the throughput gap between QA ($N_a$ = 10) and FPGAs ($N_{it}$ = 15), the dotted vertical line is the $f_{clk}^{max}$ (56 MHz) achieved by our FPGA implementation (8-bit LLR precision, fully parallel decoder), and the dark horizontal line is the upper bound of FPGA throughput ($N_{it}$=15) imposed by $f_{clk}^{max}$. The data points from top to bottom of the QBP line in the figure correspond to $N_a$ = \{1,5,10,20,50,100\} anneals respectively in all the plots.}
\label{fig:act_thrpt}
\end{subfigure}

\caption{\systemnames{} system performance in an AWGN channel. CDF in Fig.~\ref{fig:cdf_sim} is across individual LDPC problem instances. In Fig.~\ref{fig:fer_sim}, the frame size $N_F$ is bits and the Soft BP iterations are 100. In Fig.\ref{fig:act_thrpt}, all plots share common x-y axes.}
\label{fig:awgn_sys_results}
\end{figure*}

\subsection{System Performance}
\label{s:eval:system}

This section reports the \systemacronyms{} end\hyp{}to\hyp{}end
performance under the above system and design parameter choices (\S\ref{s:eval:micro}).

\subsubsection{AWGN Channel Performance}
\label{s:eval:system:awgn}

We first evaluate over a Gaussian wireless channel at SNRs in the range 
1--11~dB, comparing head-to-head against soft BP decoders operating within
various iteration count 
limits.

\paragraph{Bit error rate performance.} In Fig.~\ref{fig:ber_sim}, we investigate how average end\hyp{}to\hyp{}end BER behaves as the wireless channel SNR varies. At regions of channel SNRs less than 6~dB, \systemacronyms{} performance lags that of conventional soft BP decoders operating at 20 and 100 iterations, and differences in \systemacronyms{} performance at various QA computing times are barely distinguishable. This is because the optimal choice of $W_2$ at low SNRs is low (\S7.2.2), thus making the probability of finding the ground truth low for a QA. However as we meet SNRs greater than 6~dB, we observe \systemacronyms{} BER curves quickly drop down, reaching a BER of $10^{-8}$ at SNR 7.5--8.5~dB only, whereas conventional soft BP decoders acheive the same BER at an SNR of 10.5--11~dB. This is because the optimal choice of $W_2$ at high SNRs is high (\S7.2.2), thus separating the ground truth and the rest with a high energy gap, making the true transmitted message easier to distinguish. Our \systemacronym{} LDPC decoder acheives a performance improvement over a conventional silicon based soft BP decoder by reaching a BER of $10^{-8}$ at an SNR 2.5--3.5~dB lower.

\begin{figure*}[ht]
\centering
\begin{subfigure}[b]{0.66\linewidth}
\includegraphics[width=\linewidth]{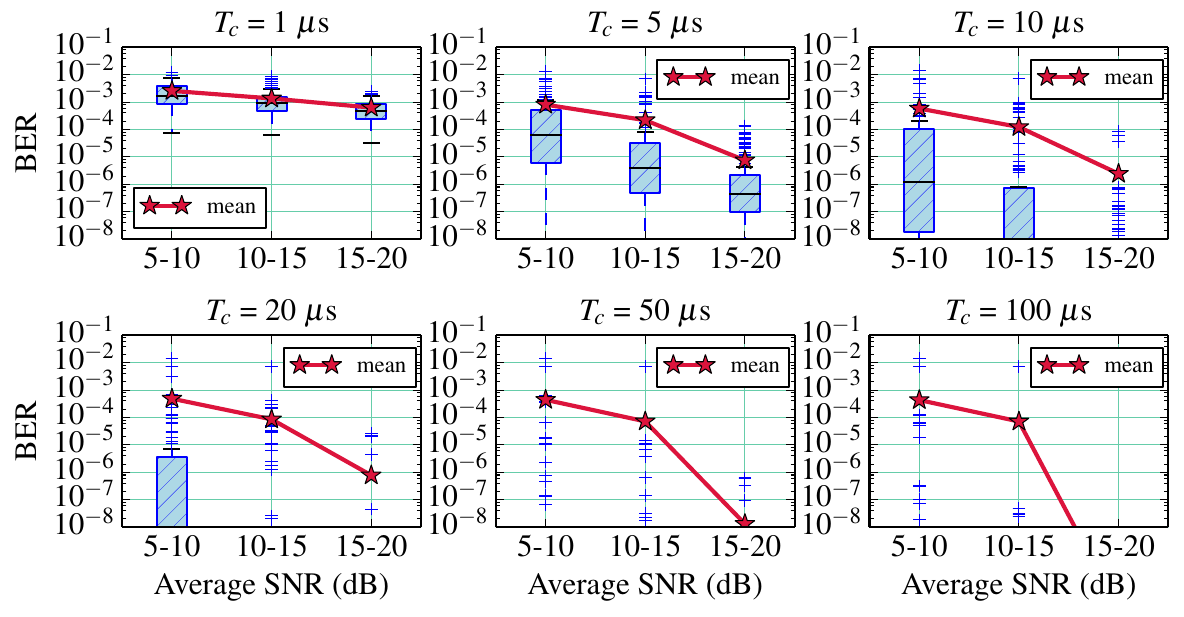}
\caption{BER of LDPC problem instances at different SNRs and QA compute times $T_c$ in trace driven channels. The missing boxes in the figure are below $10^{-8}$ BER.}
\label{fig:ber_trace}
\end{subfigure}
\hfill
\begin{subfigure}[b]{0.33\linewidth}
\includegraphics[width=\linewidth]{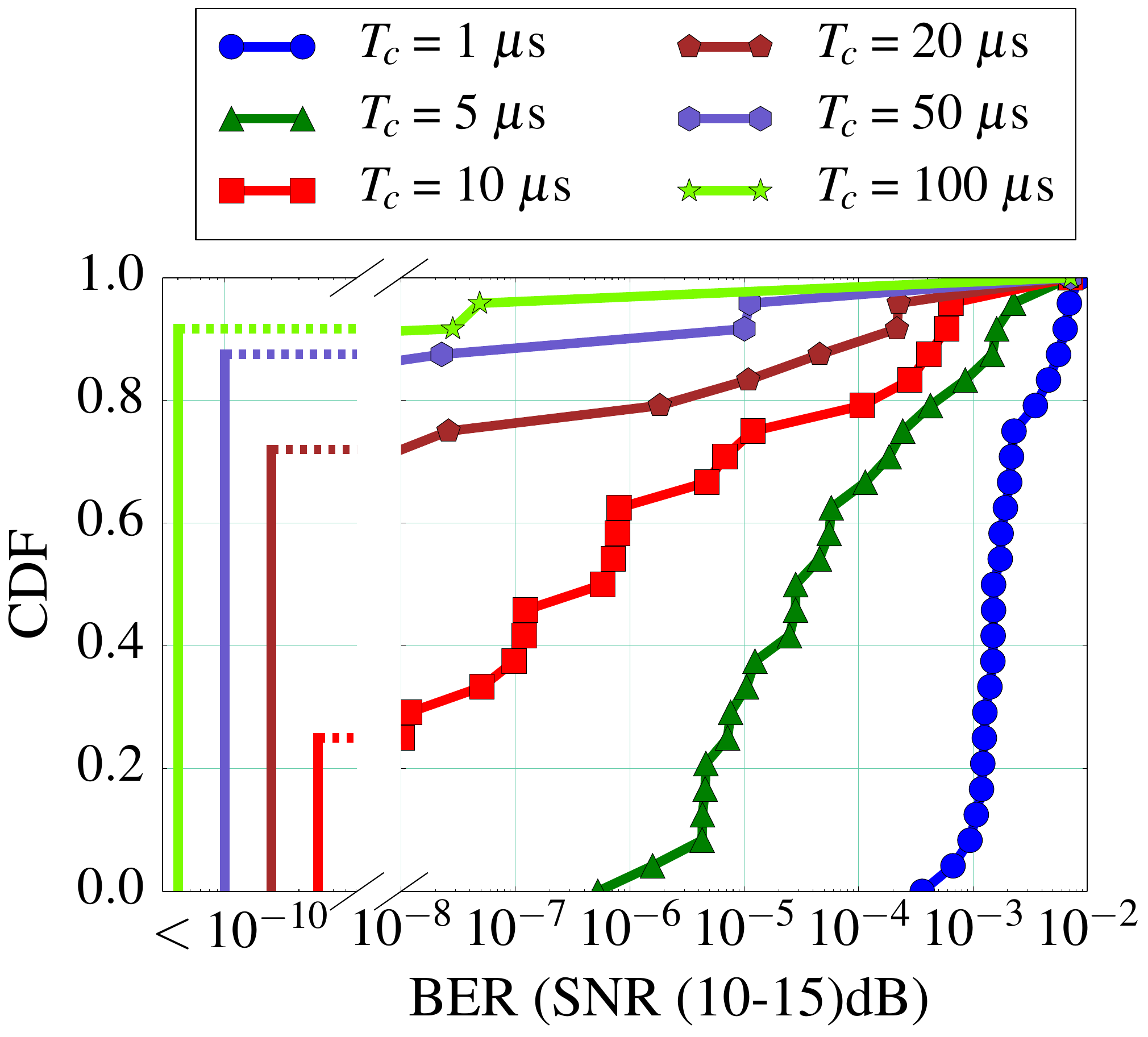}
\caption{CDF across individual LDPC problem instances in a trace driven channel.}
\label{fig:cdf_trace}
\end{subfigure}
\hfill
\caption{\systemnames{} overall experimental trace-driven channels' system performance. In Fig.~\ref{fig:ber_trace}, boxes' lower/upper whiskers and lower/upper quartiles represent $10^{th}$/$90^{th}$ and $25^{th}$/$75^{th}$ percentiles respectively.}
\label{fig:trace_sys_results}
\end{figure*}

\parahead{Across problem instances.} In Fig. \ref{fig:cdf_sim}, we investigate how bit errors are distributed among individual LDPC problem instances in the same parameter class. The figure shows that when the \systemacronym{} decoder fails due to too-low QA compute time, bit error rates are rather uniformly distributed across different problem instances. Conversely, increasing the computing time to 10--100 $\mu$s, the decoder drives BER low, so most
instances have zero bit errors, and BER variation reduces. The result shows that \{0, 28, 56, 73, 92, 98\} percent of instances under QBP's decoding are below the BER achieved by soft BP at QA compute times \{1,5,10,20,50,100\} $\mu$s respectively.

\paragraph{Frame error rate performance.} We investigate \systemacronyms{} FER performance under frame sizes $N_F$ of 420, and 12,000~bits. In Fig.~\ref{fig:fer_sim}, we observe a shallow FER error floor for SNRs less than 6 dB, noting the dependence of that error floor value on the frame length. When we meet an SNR of 8--9~dB, \systemacronym{} acheives an FER of $10^{-6}$ with low dependence on the frame length and QA compute time,
while soft BP achieves the same BER at an SNR 2--3 dB higher.

\paragraph{Throughput Analysis.} An FPGA-based LDPC decoder is bounded by a maximum operating clock frequency ($f_{clk}^{max}$), the frequency beyond which the FPGA signal routing fails. Let us define the code block \textit{solution time} $T_{st}$ as the inverse of the minimum possible time to obtain a decoded solution (\textit{i.e.,} $T_a^{-1}$ for QA and $f_{clk}\cdot N_{clk/it}^{-1}$ for an FPGA). Fig.\ref{fig:act_thrpt} reports the throughputs. The figure shows that as the channel SNR increases, the throughput gap between QA ($N_a$ = 10) and FPGAs ($N_{it}$ = 15) tends toward a constant value whose magnitude is essentially the gap between the processing throughputs of QA and FPGAs, as the value of (1--FER) \S\ref{s:fpga_throughput_eval_meth} tends toward one. The results imply that the QA can achieve a throughput improvement over the fastest FPGAs implementing a fully parallel decoder, when either the annealing time only improves roughly by 40$\times$, or when the annealing time improves by 5$\times$ in combination with a 5.4$\times$ increase of qubit resources in the QA.

Fig.\ref{fig:act_thrpt} compares QBP against soft BP for a small code of 420 bits, thus $f_{clk}^{max}$ achieved (56 MHz) is high enough for the FPGA to reach a throughput better than DW2Q QA. However, the value of $f_{clk}^{max}$ significantly reduces as code block lengths increase, due to higher complexity of the decoder. Our FPGA implementation (fully parallel decoder, 8-bit LLR precision) of a (2,3)-regular LDPC code of block length 2048 bits achieves an $f_{clk}^{max}$ of 17 MHz, while a (4,8)-regular similar LDPC code does not fit into that FPGA.

\subsubsection{Trace-driven Channel Performance}
\label{s:eval:system:traces}

Here we demonstrate \systemacronyms{} performance in real world trace driven channels (\S7.1.3).

\paragraph{Bit error rate performance.} Fig.~\ref{fig:ber_trace} depicts \systemacronyms{} BER performance in trace-driven channels. For a given compute time, we observe the BER distribution across problem instances, and its dependency on the channel SNR. For channel average-SNRs in the range 5--10~dB, we observe that a few instances lie at a high BER of $10^{-2}$, thus driving the mean BER high. As we step up to higher average SNRs greater than 10--15~dB, BER goes down very rapidly over increase in QA compute time for greater than 90\% of problem instances, since there is less probability that channel subcarriers experience very low SNRs in this scenario.

\parahead{Across problem instances.} Drilling down into individual problem instances at a particular average SNR in the range 10--15~dB, we observe in Fig. \ref{fig:cdf_trace} that more than 75\% of the problem instances lie below the $10^{-8}$ BER at computing times 20--100~$\mu$s, while exhibiting an error floor spanning two orders of BER between $10^{-4}$ and $10^{-2}$ when the QA computing time is set to 1~$\mu$s (far less than general practices).

\begin{figure}
\centering
\includegraphics[width=0.9\linewidth]{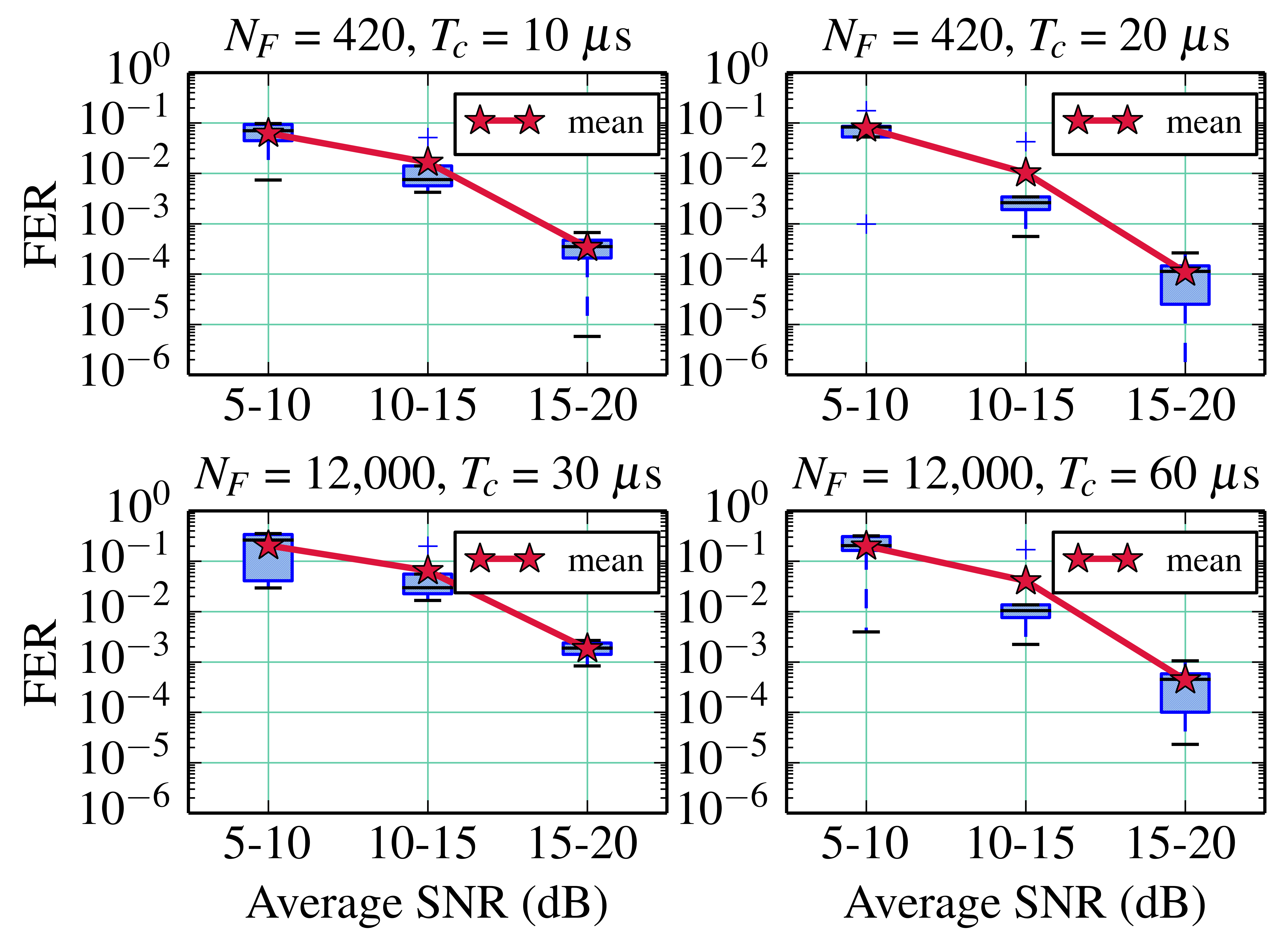}
\caption{\systemacronyms{} FER performance in trace driven channels. The unit of frame size $N_F$ in the figure is bits. In the figure, boxes' lower/upper whiskers and lower/upper quartiles represent $10^{th}$/$90^{th}$ and $25^{th}$/$75^{th}$ percentiles repectively.}
\label{fig:fer_trace}
\end{figure}

\paragraph{Frame error rate performance.} Fig.~\ref{fig:fer_trace} depicts \systemacronyms{} trace-driven channels' FER performance at various channel average-SNRs. Each box in the figure represents 10 different channel traces, where we compute FER by constructing $2\times 10^{2}$, $5\times 10^{6}$ distinct frames (as mentioned in \S7.1.2) for each channel trace when $N_F$ = 420 and 12,000 bits respectively. We observe that FER exhibits an error floor when the average channel SNRs are less than 10--15~dB. FER drastically drops down for channel SNRs greater than 15~dB.

%% file: related.tex
\section{Related Work}
\label{s:related}

Bian \textit{et al.}\ \cite{bian2014discrete} present discrete optimization problem solving techniques tailored to QA, solving the LDPC decoding problem by dividing the Tanner graph into several \textit{sub\hyp{}regions} using min\hyp{}cut heuristics, where a different QA run solves each sub\hyp{}region. Bian \textit{et al.}\ coordinate solutions of each run to construct the final decoded message. Conversely, \systemacronyms{} approach differs with \cite{bian2014discrete} both with respect to QUBO formulation and QA hardware embedding. The Bian \emph{et al.}\ QUBO design does not adapt to both the wireless channel noise (distance function \S\ref{s:dist_fcn}) and the binary encoding minimization of the ancillary qubits (LDPC satisfier function \S\ref{s:sat_fcn}). From embedding perspective, \systemacronym{} can solve up to 280 check constraints in a single anneal while Bian \textit{et al.}\ solves up to only 20 check constraints on an earlier QA with 512 qubits (which extends to 60--80 check constraints on the current QA with 2,048 qubits). Bian \textit{et al.}\ evaluate over a binary symmetric channel (each sub-region run with $T_a = 20\mu s$) with crossover probabilities in the range of 8--14\%, unrealistically high for practical wireless networks, nonetheless experiencing that only 4\% out of $10^4$ anneals had no bit errors, lower\hyp{}bounding their BER by $10^{-3}$. Lackey proposes techniques for solving Generalized BP problems by sampling a Boltzmann distribution \cite{lackey2018belief}, but does not venture into a performance evaluation. It is also possible to use the \systemacronyms{} QUBO design (\S\ref{sec:5.1}) as an input to D-Wave's built\hyp{}in greedy search embedding tool \cite{cai2014practical}, but this approach scales up to only 60 (2,3)-regular LDPC check constraints, which limits the LDPC code block length to an impractical 90 encoded bits.

QA machines have been recently used to successfully optimize problems in several adjacent domains including Networks \cite{wang2016quantum, 10.1145/3341302.3342072}, Machine Learning \cite{mott2017solving, adachi2015application}, Scheduling \cite{venturelli2015quantum}, Fault Diagnosis \cite{bian2016mapping, perdomo2015quantum}, and Chemistry \cite{10.1007/978-3-030-14082-3_10}. Efficient embedding methods for mapping fully-connected QUBOs on to QA hardware graphs have also been discussed \cite{PhysRevX.5.031040, boothby2016fast} which support up to 64 variables on DW2Q QA.

%% file: discussion.tex
\section{Looking Forward}
\label{s:discussion}

\begin{figure}
\centering
\includegraphics[width=0.86\linewidth]{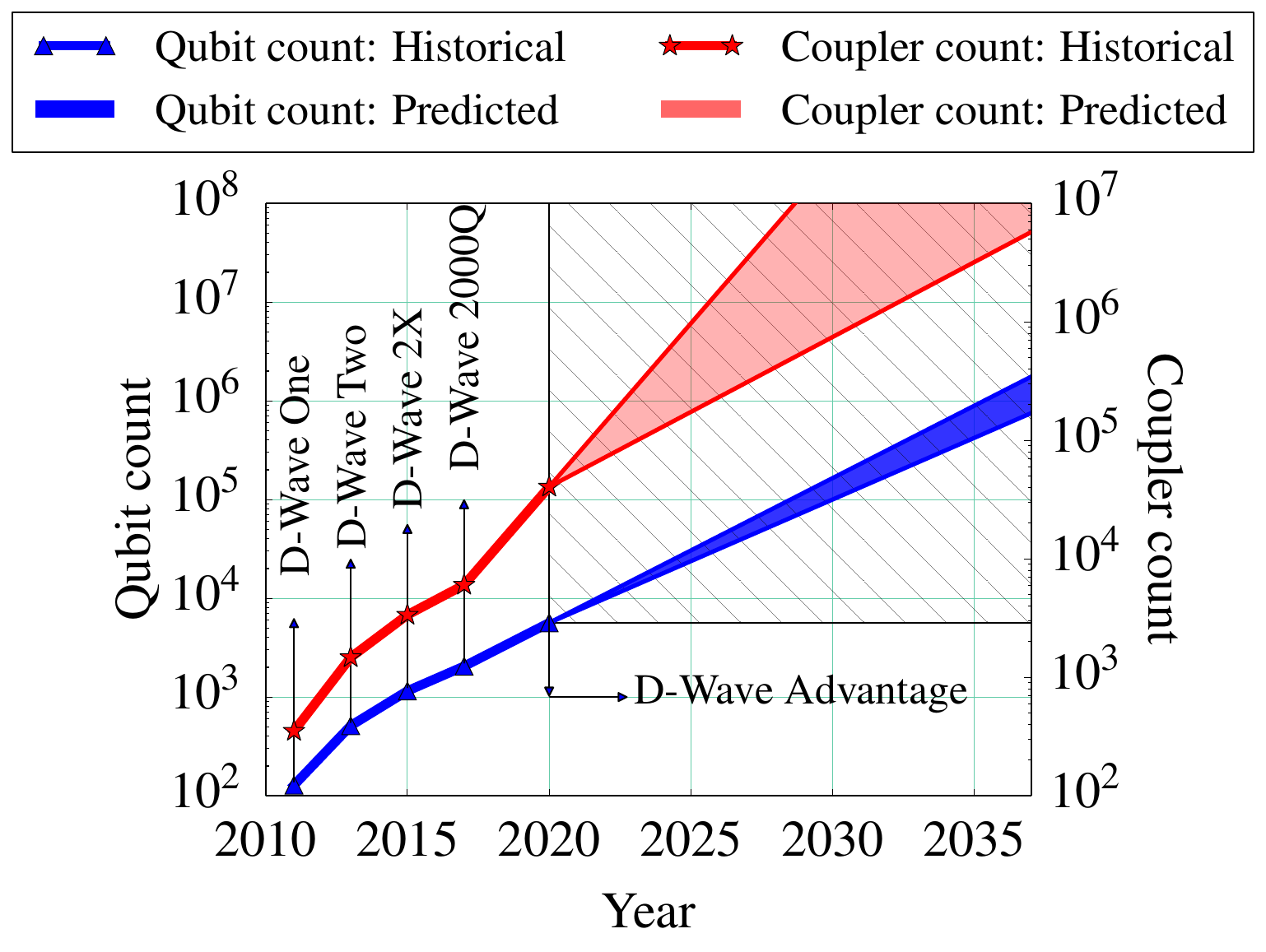}
\caption{D-Wave QA's hardware resource counts over time. Historical data is in the years 2011--2020. The blue filled (darker) and the red filled (lighter) areas are the predicted qubit and coupler counts respectively, whose upper/lower boundaries are extrapolations of the most recent 2017--2020/2015--2017 qubit-coupler growths respectively. Annotations in the figure are the QA processor titles in the respective years.}
\label{fig:extrapolation}
\end{figure}

\parahead{QA hardware trend predictions.} For the past decade, the number of physical qubits in D-Wave's QPU has been steadily doubling each year and this trend is expected to continue \cite{dwave}. Fig.~\ref{fig:extrapolation} presents a predicted extrapolation of quantum annealer qubit and coupler counts into the future. The figure shows that at these rates, an annealer processing chip with one million qubits could be available roughly by the year 2037. Let us envision future QAs with a processor topology that is either a Chimera or a supergraph of Chimera (\textit{e.g.,} Pegasus \cite{pegasus}) with $N_Q$ available qubits, which enables QBP to decode block lengths of at most 5$N_Q$/24~bits in a single anneal. Thus in a QA with $N_Q$ = \{$10^4$, $10^5$, $10^6$\} qubits, we forecast QBP to be able to decode LDPC codes of block lengths up to \{2,083, 20,833, 208,333\} bits respectively in a single anneal with peak processing throughputs reaching \{0.694, 6.94, 69.4\} Gbps respectively, while most classical fully parallel decoders do not implement block lengths exceeding 2,048 bits due to signal routing and clock frequency constraints \cite{hailes2016survey}.

\parahead{Limitations of QA.} The lack of all-to-all qubit connectivity in today's QPUs limits the size of the problems the QA can practically solve, implying that the requirement of embedding is a major impediment to leveraging QA for practical applications. Furthermore, the process of transferring the computation and running on real analog QA device introduces a source of noise distinct from communication channel noise called \textit{intrinsic control error} or \textit{ICE}, which arises due to the flux noise and the quantization effects of the qubits. ICE effects in the QA alter both the problem biases $(h_i \to h_i \pm \delta h_i)$ and couplers $(J_{ij} \to J_{ij} \pm \delta J_{ij})$, leading the QA to solve a slightly modified input problem in each anneal. Although the errors $\delta h_i$ and $\delta J_{ij}$ are currently in the order of $10^{-2}$, they may degrade the solution quality of some problems whose minimum energy state is not sufficiently separated from the other states in the energy landscape of the input problem Hamiltonian \cite{dwavetech}. From a design perspective, ideally the qubits that are embedded together must all agree and end up in a similar final state at the end of the annealing process, otherwise the embedding chain is said to be \textit{broken}: typically broken chains lead to performance degradation, and they are more likely to occur when the number of qubits embedded together in a particular chain is large $(> 10)$. \systemacronyms{} embedding design include chains of length two--four, and five--nine for Level~I and Level~II embeddings (\S\ref{s:design:embedding}) respectively. Further, there exist post\hyp{}processing techniques such as \textit{majority vote}, \textit{weighted random}, and \textit{minimize local energy} that can be used to improve the performance of broken chains \cite{qbsolv}. QBP's embedding results in a low fraction of broken chains ($\approx$ 2\%), and we use the majority voting technique in those cases to find the variables involved.

\begin{figure}
\centering
\includegraphics[width=0.75\linewidth]{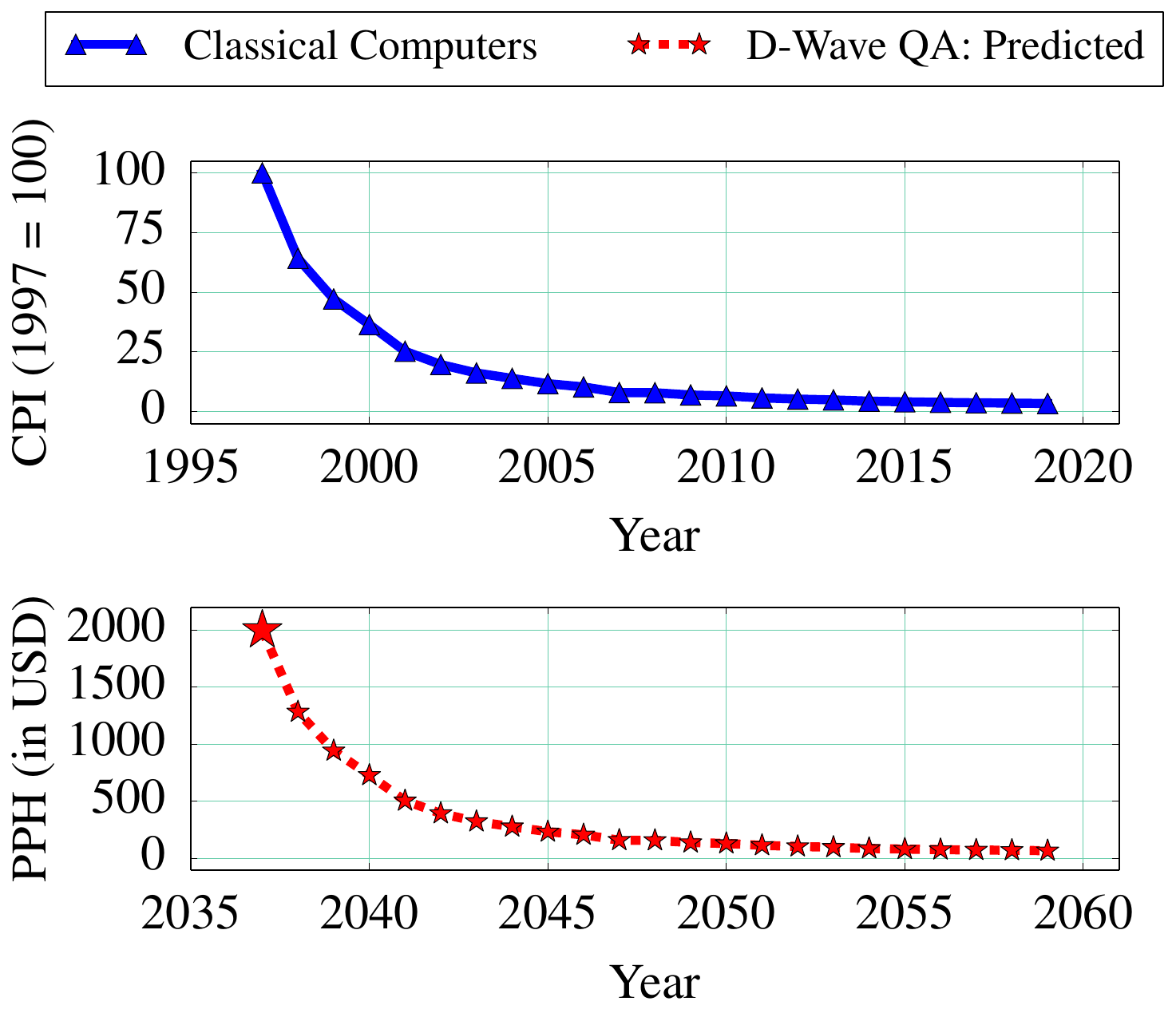}
\caption{\textit{Top.} The plot shows the consumer price index (CPI) of classical computers and peripherals over time with 1997 as the base year. \textit{Bottom.} The plot shows the predicted \emph{price per hour} (PPH) of quantum annealers over time. The larger data point is the actual 2015--2020 QA price, which is conservatively assumed to remain the same until the QA technology matures in a predicted 17 years.}
\label{fig:cost}
\end{figure}

\parahead{Cost considerations.} QA technology is currently a cloud-based system and currently costs USD \$2,000 for an hour of QPU access time, which is approximately \$17.5M for an year. As the evolution of the technology is currently at an early stage (2011--), we consider the next 15 years for the technology to mature to the market. As usage becomes more widespread in future years, we hypothesize that QA prices will decrease with the same trend as classical compute prices have done since the late 20th century. Fig.~\ref{fig:cost} (top) shows the consumer price index (CPI) of classical computers and peripherals over time \cite{cost}, while Fig.~\ref{fig:cost} (bottom) shows a similar predicted trend for QA \textit{price per hour} (PPH). Figs.~\ref{fig:extrapolation} and \ref{fig:cost} imply that, at these rates QA technology is expected to deliver a machine with more than $10^6$ qubits on a single annealer processing chip at the prices of \$730, \$235, \$130, \$82, and \$68 per hour of QPU access time, by the years 2040, 2045, 2050, 2055, and 2059, respectively. This represents an approximate projected cost of \$6.4M, \$2M, \$1.1M, \$700K, and \$600K per year, by the above respective years.

\parahead{Timing considerations.} Currently, the DW2Q has a 30--50~ms preprocessing time, 6--8~ms programming time, and 0.125~ms solution readout time per anneal, which are beyond the processing times available for wireless technologies (3-10~ms) \cite{10.1145/3341302.3342072}, with supported annealing times in the range [1~$\mu$s, 2~ms]. Given the large amount of cost, embedding, and timing overheads of today's annealers, QBP currently cannot be deployed for use in practical applications. While approaches \cite{bian2014discrete, qbsolv} that decompose large\hyp{}scale optimization problems can be used to study more problem variables, they suffer from requiring additional factors of the aforementioned machine overhead times for each extra anneal. The historical trend is encouraging, with the DW2Q having a 5$\times$ annealing time improvement over the circa\hyp{}2011 D-Wave One \cite{boixo2014evidence}.

%% file: conclusion.tex
\section{Conclusion and Future Work}
\label{s:concl}

\systemacronym{} is a novel QA\hyp{}based uplink LDPC decoder that makes efficient use of the entire QA hardware to achieve new levels of performance beyond state\hyp{}of\hyp{}the\hyp{}art BP decoders. Further efforts are needed to generalize QBP's graph embedding to large-scale LDPC codes with higher check bit degrees. The techniques we propose here may in the more distant future come to be relevant to practical protocol settings, while application of the aforementioned Cloud/Centralized-RAN architecture has
also been proposed for small cells \cite{fluidnet-ton16,10.5555/3001647.3001667}: opening the possibility to its future application
to managed Wi-Fi local\hyp{}area networks. Investigating the QA technology for problems such as network security, downlink precoding, scheduling, and other uplink channel codes such as Polar and Turbo codes is potential future work direction.

%% file: acknowledgements.tex
\section*{Acknowledgements}

We thank the anonymous shepherd and reviewers of this paper for their extensive technical feedback, which has enabled us to significantly improve the work. We also thank Davide Venturelli, Catherine McGeoch, the NASA Quantum AI Laboratory (QuAIL), D\hyp{}Wave Systems, and the Princeton Advanced Wireless Systems (PAWS) Group for useful discussions. This research is supported by National Science Foundation (NSF) Award CNS\hyp{}1824357, a gift from InterDigital corporation, and an award from the Princeton University School of Engineering and Applied Science Innovation Fund.  Support from the USRA Cycle 3 Research Opportunity Program allowed machine time on a D-Wave machine hosted at NASA Ames Research Center.